\documentclass[aps,amsmath,amssymb,footinbib,12pt]{revtex4}

\usepackage{amsmath,amssymb,stmaryrd}   
\usepackage[T1]{fontenc}
\usepackage[latin9]{inputenc}
\usepackage{array}
\usepackage{float}
\usepackage{amstext}
\usepackage{amssymb}
\usepackage{graphicx}
\usepackage{amsmath}
\usepackage{enumitem}
\usepackage{natbib}
\usepackage{float}
\usepackage{verbatim}
\usepackage{mathrsfs} 
\usepackage{color}
\usepackage{bm}

\newcommand{\beq}{\begin{equation}}
\newcommand{\eeq}{\end{equation}}
\newcommand{\eeqr}{\end{eqnarray}}
\newcommand{\beqr}{\begin{eqnarray}}
\newcommand{\gd}{\Gamma_{\downarrow}}
\newcommand{\gu}{\Gamma_{\uparrow}}

\begin{document}
\title{Entropy production in quantum  is different}
\author{Mohammad H. Ansari}
\email{m.ansari@fz-juelich.de}
\affiliation{J\"ulich-Aachen Research Alliance Institute (JARA) and  Peter Gr\"unberg Institute (PGI-2), Forschungszentrum J\"ulich, D-52425 J\"ulich, Germany}

\author{Yuli V. Nazarov}
\affiliation{Department of Quantum Nanoscience, Kavli Institute of Nanoscience, TU Delft, Lorentzweg 1, 2628CJ Delft, The Netherlands}

\author{Alwin van Steensel}
\affiliation{J\"ulich-Aachen Research Alliance Institute (JARA) and  Peter Gr\"unberg Institute (PGI-2), Forschungszentrum J\"ulich, D-52425 J\"ulich, Germany}

\begin{abstract}
Currently, `time' does not play any essential role in quantum information theory. In~this sense, quantum information theory is underdeveloped similarly to how quantum physics was underdeveloped before Erwin Schr\"odinger introduced his famous equation for the evolution of a quantum wave function.   In this review article, we cope with the problem of time for one of the central quantities in quantum information theory: entropy.  Recently, a replica trick formalism, the so-called `multiple parallel world' formalism, has been proposed that revolutionizes entropy evaluation for quantum systems. This formalism is one of the first attempts to introduce `time' in quantum information theory. With the total entropy being conserved in a closed system, entropy can flow internally between subsystems; however, we show that this flow is not limited only to physical correlations as the literature suggest. The  nonlinear dependence of entropy on the density matrix introduces new types of correlations with no analogue in physical quantities. Evolving a number of replicas simultaneously makes it possible for them to exchange particles between different replicas. We~will summarize some of the recent news about entropy in some example quantum devices. Moreover, we take a quick look at a new correspondence that was recently proposed that provides an interesting link between quantum information theory and quantum physics. The mere existence of such a correspondence allows for exploring new physical phenomena as the result of controlling entanglement in a quantum device.
\end{abstract}

\maketitle
\tableofcontents{}

\section{Introduction}
Entropy is one of the central quantities in thermodynamics and, without its precise evaluation, one cannot predict what new phenomena are to be expected in the thermodynamics of a device. In~quantum theory, entropy is defined as a nonlinear function of  the density matrix, i.e.,~$S=-\textup{Tr}\hat{\rho}\ln\hat{\rho}$, in~the units of the Boltzmann constant $k_B$. The~mere nonlinearity indicates that entropy is not physically observable because, by~definition,  observables are linear in the density matrix. Let us further describe this statement. Here, we do not assume that the density matrix is a physical quantity. The~reason is that evaluating all components of a many-body density matrix requires many repetitions of the same experiment with the same initial state. Not only is this difficult but also the fact that measurement changes quantum states prevents exact evaluation. A~physical quantity, such as energy or charge, can be measured in the lab in real time and can be defined in quantum theory to linearly depend on the density matrix. This is not true for entropy and therefore we cannot assume it is a physical quantity directly measurable in the~lab.

In fact, the~precise time evolution of entropy is still an open problem and has not been properly addressed in the literature ~\cite{{Pol 2011},{Santos 2011},{QuantumInformation}}. A~consistent theory of quantum thermodynamics can only be achieved after finding nontrivial relations between the quantum of information and physics. In~recent years, exquisite mesoscopic scale control over quantum states has led technology to the quantum realm. This~has motivated exploring new phenomena such as exponential speed up in computation as well as power extraction from quantum coherence~\cite{{qcom},{Lagoudakis},{ANSexact},{qcom2},{AnFr}}. Recently, there have been attempts to implement quantum versions of heat engines using superconducting qubits~\cite{pekola}. However, recent developments in realizing quantum heat engines, such as in References~\cite{{uzdin},{schul},{pop}}, rely on semiclassical stochastic entropy production after discretizing energy. A~long-lasting question is how the superposition of states transfers heat and how much entropy is produced as the result of such a~transfer.  

A quantum heat engine (QHE) is a system with several discrete quantum states and, similar to a common heat engine, is connected to several environments kept at different temperatures. In~fact, a~number of large heat baths in these engines share some degrees of freedom quantum mechanically. Such a system is supposed to transfer heat according to the laws of quantum mechanics. The~motivation
for research in QHE originates from differences they may controllably make on the efficiency and output powers. Let us consider the example of two heat baths $A$ and $B$, both coupled through a quantum system $q$ that contains  discrete energies  and allows for the superposition of states with long coherence time. Let us clarify that, in~this paper, we study the flow of thermodynamic Renyi and von Neumann entropies between the heat baths and quantum system $q$. Therefore, other entropies are beyond the scope of this paper. 
This quantum system coupled to the two large heat baths is in fact a physical quantum system that is energetically coupled to the reservoirs and allows for stationary flow of heat as well as a flow of thermodynamic entropy from one reservoir to another.  We will see in the next section that, similar to physical quantities such as energy and charge, the~total entropy of a closed system is a conserved quantity and does not change in time. However, internally, entropy can flow from one subsystem to another. Therefore, sub-entropies may change in time and this change may indicate a change in the energy transfer.  Some important questions one may ask are:  \emph{Does a quantum superposition change entropy?} 
This is one of the questions that we will address in this almost pedagogical review paper and we will furthermore describe how the information content in entropy can be meaningful in~physics. 

In a typical engine made of reservoirs $A$, $B$ and an intermediate quantum system $q$ with discrete energy levels, the~change of entropy in one of the reservoirs, say $B$, between~the time $0$ and $t$ is $S_{B} \left( t \right) - S_{B} \left( 0 \right) = -\textup{Tr}\left\{ \rho\left(t\right)\ln\rho_{B}\left(t\right)\right\} -\textup{Tr}_{B}\left\{ \rho_{B}^{\textup{eq}}\ln\rho_{B}^{\textup{eq}}\right\} $, where in the first term we have safely replaced one of the two partial density matrices with the total density matrix, and~accordingly replaced the partial trace with total one. The~conservation of entropy tells us that the total entropy maintains its initial value at the separable compound state $\rho\left(0\right)=\rho_{q}\left(0\right)\rho_{A}^{\textup{eq}}\rho_{B}^{\textup{eq}}$,
i.e., $ -\textup{Tr}\left\{ \rho\left(t\right)\ln\rho\left(t\right)\right\} =-\textup{Tr}_{q}\left\{ \rho_{q}\left(0\right)\ln\rho_{q}\left(0\right)\right\} -\sum_{i=A,B}\textup{Tr}_{i}\left\{ \rho_{i}^{\textup{eq}}\ln\rho_{i}^{\textup{eq}}\right\}$.  After~a few lines of algebra one can find that the change of entropy at the reservoir  is $S_{B}\left(t\right)-S_{B}\left(0\right)=S_{B}\left(\rho\left(t\right)||\rho_{A}^{\textup{eq}} \rho_{B}\left(t\right)\rho_{q}\left(0\right)\right)+\sum_{i=q,A}\textup{Tr}_{i}\left\{ \left(\rho_{i}\left(t\right)-\rho_{i}\left(0\right)\right)\ln\rho_{i}\left(0\right)\right\} $,
with $S\left(\rho||\rho'\right)\equiv\textup{Tr}\left\{ \rho\ln\rho\right\} -\textup{Tr}\left\{ \rho\ln\rho'\right\} $ being the relative entropy. 
Since relative entropy is a positive number~\cite{Lieb} and equals zero only
for identical density matrices $\rho=\rho'$, the~first part of the entropy flow is positive and irreversible. This satisfies the classical laws of thermodynamics. We will show that, in contrast to what has been so far presented in the literature~\cite{Esposito11}, the~second term in the entropy flow is {\it not} heat transfer---the average change of energy at the two times $Q_{B}\equiv\langle H\left(0\right)\rangle_{B}-\langle H\left(t\right)\rangle_{B}$. Instead, it is the difference of incoherent and coherent heat transfers~\cite{AN15}, i.e.,~$\left( Q_{B,\textup{incoh}} \left( t \right) - Q_{B,\textup{coh}} \left(t\right) \right) - \left(Q_{B,\textup{incoh}} \left(0\right) - Q_{B,\textup{coh}} \left(0\right) \right)$. This is the new result that heavily modifies the flow of entropy in some quantum heat engines and leads to some recent new physics~\cite{{ANS16},{Schully2018},{Utsumi19-1},{Utsumi19-2}}. 

In this review paper, we look at some of the simplest and most important quantum heat engines. Depending on the external drive or internal degeneracy, the~exact evaluation of entropy is indeed very different from what has been presented in the literature so far. We will describe how to precisely evaluate entropy and its flow by using a replica trick that properly allows for the mathematically involved nonlinearity. We introduce a new class of correlations that allow information transfers and are different from physical correlations. For~equilibrium systems, these informational correlations satisfy a generalized form of Kubo--Martin--Schwinger (KMS) relation~\cite{{kms1},{kms2}}. This part of the analysis will be presented in a self-contained fashion after reviewing some of the classical and quantum definitions of entropy and introducing our replica trick for evaluating the time evolution of generalized Keldysh contours. We describe a short protocol for evaluating Keldysh diagrams and in some examples perform the evaluation of a number of diagrams. We present results of example quantum devices such as a two-level quantum heat engine, a~photocell,
as well as a resonator, each one mediating heat transfer between two large heat baths. Finally, we briefly report on the new correspondence that makes entropy flow directly measurably in the lab by monitoring physical quantities, i.e.,~the statistics of energy~transfer.
 

\section{Classical~Systems}
\unskip

\subsection{Classical~Entropy}

Many systems in classical physics carry entropy. Some of the most studied systems are: charge transport at a point contact~\cite{{Levitov},{LevitovKlysh}}, energy transport in heat engines~\cite{Kindermann}, and~a gravitational hypersurface falling into a black hole~\cite{{blackhole},{ANSb1},{ANSb2},{ANSb3}}. Let us for simplicity of the discussion review classical entropy by means of the example of charge transport through a point contact.  Consider for this purpose two large conductor plates connected at a point, the~so-called `point contact system'. This~classical point contact either transmits a charged particle with probability $p$ or blocks the  transmission with probability $1-p$. Let us consider $N$ attempts take place. For~$N\gg 1$ it is most likely that, in $pN$ out of $N$ times, the particles are successfully transferred and, in $\left(1-p\right)N$ out of $N$ times, they are not. For~unmarked particles, the order of events does not matter, therefore the number of possibilities with $pN$ transfers out of $N$ attempts is
\begin{equation} \label{eq. class comb}
\mathcal{N}=\left(\begin{array}{c}
N\\
pN
\end{array}\right)\approx\frac{N^{N}}{\left(pN\right)^{pN}\left[\left(1-p\right)N\right]^{\left(1-p\right)N}}=\frac{1}{p^{pN}\left(1-p\right)^{\left(1-p\right)N}}.
\end{equation}

This number rapidly grows with $N$. In~order to keep the number small, we take its logarithm. This~defines the so-called Shannon
entropy, i.e.,~$S_{Shannon}=\log_{2}\mathcal{N}=-N\left[p\log_{2}p+\left(1-p\right)\log_{2}\left(1-p\right)\right]$.

The linear dependence of the Shannon entropy on the number of attempts $N$ indicates its additivity. The~definition of entropy can be generalized to account for extended geometries such as a $k$ + 1-path terminal that connects any reservoir to $k$ others. In~this case, $k$ probabilities contribute to understanding the possibility of transmission from a reservoir to any one of the other $k$ reservoirs, thus entropy is generalized to  $S_{Shannon}=-N\sum_{n=1}^kp_{n}\log_{2}p_{n}$. This entropy may vary in time.  One possible reason for such variation could be due to  time-dependent probabilities $p_n(t)$. Another possibility for time evolution of entropy could be the presence of some bias in controlling the system. For~example, consider that, after one successful transfer, the transmission is reduced or closed for a rather long time before it opens again to another transfer attempt. The~entropy of such a system depends on whether or not a success transfer has taken place in the~past. 

In fact, in this paper, what we call entropy production refers to the time variation of partial entropy associated with a part of a closed system. Moreover, as~stated in the Introduction, in~this paper, we are only interested in the time variation in thermodynamic systems such as heat baths; therefore, our focus is only on thermodynamic entropies and its time evolution, namely `entropy production'. In~this section, although we discuss Shannon entropy $S_{Shannon}$, we have to  distinguish between the Shannon entropy, which can be measured as a number of bits, and~the rest of the paper in which we study von Neumann thermodynamic entropy measured in the unit Joule per Kelvin. The~Shannon entropy and the thermodynamic entropy are related by the Boltzmann constant $k_B$, i.e.,~$ S_{Thermodynamic}=k_B S_{Shannon}$.  Without~the loss of generality, we use the convention that  $k_B=1$, although~the reader should keep in mind that, in this paper, we are interested in finding changes in thermodynamic entropy flow as the result of energy exchange~processes.

\subsection{Renyi~Entropy}

Alfred Renyi introduced the generalization of Shannon entropy that maintains the additive property~\cite{renyi}. For~a finite set of $k$ probabilities $p_{i}$ with $i=1,\cdots,k$, the Renyi entropy of degree $M$ is defined as
\begin{equation} \label{eq. stan Renyi}
\mathcal{S}_{M}=\frac{1}{1-M}\log\sum_{i}\left(p_{i}\right)^{M},
\end{equation}
with positive entropy order $M>0$. The~symbol $\mathcal{S}_M$ indicates that this is the original definition of Renyi entropy to make it distinct from the simplified definition $S_M$ we use in this paper. The~constant prefactor $1/(1-M)$ in Equation~(\ref{eq. stan Renyi})  has certain advantages. One of the advantages is that it helps to compactify the definition of some other entropies using Equation~(\ref{eq. stan Renyi}); i.e.,~the analytical continuation of Renyi entropy in the limit of $M$ approaching 1 ($\infty$) defines Shannon (min) entropy. Another advantage of the prefactor is that it allows for interpretation of the quantity as the number of bits (thanks to one of the referees for pointing out these remarks).

Here, we present a simplified version of the definition. The~logic behind such simplification is that the calculation in the limits requires  L'Hopital's rule; i.e.,~ $S_{Shannon,\ min}=\lim_{M\to 1,\infty} S_{M}^{\textup{R}}=-\lim_{M\to 1,\infty} d (\log\sum_{i}\left(p_{i}\right)^{M}) /dM$. We define a rescaled  Renyi entropy, which is different from the original definition by a prefactor $1/(M-1)$:
\begin{equation} \label{eq. Renyi}
S_{M}=-\log\sum_{i}\left(p_{i}\right)^{M}.
\end{equation}

The reason to define the simplified formula is that evaluating entropy itself is beyond the scope of this paper. Instead, we need to find the time derivative of the entropy (i.e., entropy flow). Due to the presence of a logarithm in Equation~(\ref{eq. Renyi}), any contact prefactor in the definition of entropy will be canceled out from the numerator and denominator of entropy flow.  The~only trouble is that we must keep in mind that the Shannon entropy can be reproduced after taking the $dS_M/dM$ in the limit of $M \to 1$. In~fact, given that $dx^{M} / dM = d\exp\left(M\ln x\right)/dM=x^{M}\ln x$, one can write
\begin{equation}
\lim_{M\to1}\frac{dS_{M}}{dM}  =  -\lim_{M\to1}\frac{\sum_{i}\left(p_{i}\right)^{M}\ln p_{i}}{\sum_{i}\left(p_{i}\right)^{M}}=-\sum_{i}p_{i}\ln p_{i}=S_{Shannon}.
\end{equation}

 In the rest of the paper, we use the simplified definition. However, given that the difference between the two definitions is marginal, only a constant factor, the reader may decide to use either definition, subject to the discussion~above.

In a point contact, given that Renyi entropy is additive  for independent attempts, the~total Renyi entropy after $N$ uncorrelated attempts will be $S_{M} = -N \log \left( p^{M} + (1-p)^{M} \right)$. In~a classical heat reservoir, the Renyi entropy is more closely related to free energy. Consider a bath at temperature $T$ with a large number of energy states $\epsilon_{i}$. The~corresponding Gibbs probabilities are $p_{i} = \exp \left( -\epsilon_i T \right) / Z (T)$ and $Z\left(T\right)\equiv\sum_{i}p_{i}$ is the corresponding partition function. The~Renyi entropy of the heat bath is $S_{M}=-\ln\left(\sum_{i}\exp\left(-M\epsilon_{i} T\right)\right)+M\ln Z\left(T\right)$. The~free energy will be $F\left(T\right)=-T\ln Z\left(T\right)$, which is related to the Renyi entropy as $S_{M} = \left( M / T \right) \left( F \left( T \right) - F \left( T / M \right) \right)$, i.e.,~the free energy difference at temperatures $T$ and $T/M$. 

\section{Quantum}
\unskip

\subsection{Von Neumann and Renyi~Entropy}

Let us now consider that a large system $A$ with many degrees of freedom interacts with a small quantum system $q$.  This can be thought of as the two share some degrees of freedom.  The~two exchange some energy via those shared degrees of freedom. Quantumness indicates that $q$ carries a discrete energy spectrum and can be found in superposition between energy levels. Let $\rho$ be the density matrix of the compound system.  The~partial density matrix of  $A$ is defined by tracing out the system $q$ from $\rho$, i.e.,~$\rho_{A}=\textrm{Tr}_{q}\rho$. The~von Neumann entropy for system $A$ in the Boltzmann constant unit is defined as
\begin{equation} \label{eq. vonNeumann}
S^{\left(A\right)}=-\textrm{Tr}_{A}\rho_{A}\ln\rho_{A}
\end{equation}
and the generalization of entropy in quantum theory will naturally give rise to defining the following quantum Renyi entropy for system $A$:
\begin{equation} \label{eq. q.renyi}
S_{M}^{\left(A\right)}=-\ln\textrm{Tr}_{A}\left(\rho_{A}\right)^{M}.
\end{equation}

The density matrix of the isolated compound system evolves between the times $t'$ and  $t>t'$ using a unitary transformation that depends on the time difference $U\left(t-t'\right)$. Therefore, one can evaluate $\textrm{Tr}_{A}\left(\rho_{A}\right)^{M}$  using the unitary transformation to trace it back to the time $t'$; i.e.,~  
\begin{eqnarray*} \label{eq. unitev}
\textrm{Tr}\left(\rho\left(t\right)\right)^{M} & = & \textrm{Tr}\left\{ \left(U\left(t-t'\right)\rho\left(t'\right)U^{\dagger}\left(t-t'\right)\right)^{M}\right\} =\textrm{Tr}\left\{ U\left(t-t'\right)\rho\left(t'\right)^{M}U^{\dagger}\left(t-t'\right)\right\} \\
 & = & \textrm{Tr}\rho\left(t'\right)^{M}.
\end{eqnarray*}

After taking the logarithm from both sides, one finds that the Renyi entropy remains unchanged between the two times $t$ and $t'$. In~other words, in a closed system, similar to energy and charge, Renyi entropy is a conserved quantity:
\begin{equation} \label{eq. conserv S}
\frac{dS_{M}}{dt}=0.
\end{equation}

Let us consider for now that there is no interaction between  $A$ and $q$. One can expect naturally that partial entropies are conserved as the result of no interaction because~each subsystem can evolve with an independent unitary operator:
\begin{equation} \label{eq. par ent flow}
\frac{dS_{M}^{\left(A\right)}}{dt}=\frac{dS_{M}^{\left(q\right)}}{dt}=0.
\end{equation} 

Interesting physical systems interact. Therefore, let us now consider that $A$ and $q$ interact. Consider that the total Hamiltonian is $H=H_{A}+H_{q}+H_{Aq}$. For~interacting systems, there is an important difference between conserved physical and information quantities. For~physical quantities, the~conservation holds in the whole system as well as in each
subsystem. As~far as Renyi entropies are concerned, there is a conservation law for the total Renyi entropy $\ln S_M^{(A+q)}$; however, this quantity is only approximately equal to the sum $\ln S_M^{(A)} + \ln S_M^{(q)}$, up~to the terms proportional to the volume of the system. Therefore, no exact conservation law can be expected  for the extensive quantity summation: $\ln S_M^{(A)} + \ln S_M^{(q)}$ \cite{deu}.  The reason is that, although the evolution of the entire system is governed by a unitary operator, the subsystem evolves non-unitarily.  In~the limit of weak coupling $|H_{Aq}|/|H_{A}+H_{q} | \ll 1$, the entropy of entire system can only be approximated with the sum of two partial entropies, thus the sum of partial entropies can only approximately satisfy a conservation, i.e.,~ $dS_{M}^{\left(A\right)} / dt + dS_{M}^{\left(q\right)} / dt \approx 0$. Outside of the validity of the weak coupling approximation, we must expect
that, although~the total entropy conserves, the interacting parts have entropy flows different from each other:
\begin{equation} \label{eq. noneq par flow}
\frac{dS_{M}^{\left(A\right)}}{dt}\neq-\frac{dS_{M}^{\left(q\right)}}{dt}.
\end{equation}

This makes the conservation of Renyi entropy different from the conservation of physical quantities. The~root for the difference is in fact in the nonlinear dependence on the density matrix, namely `non-observability' of entropy~\cite{Nazarov11}.

\subsection{Replica~Trick}

Calculating the full reduced density matrix for a general system is the subject of active research. Here, we use a different method that
is reminiscent of the `replica trick' in disorder systems. The~trick has been introduced in the context of quantum field theory by Wilczek~\cite{Wilczek} and Cardy~\cite{Cardy} and later in the context of quantum transport by Nazarov~\cite{Nazarov11}. The~key point is that, if we can evaluate $\textrm{Tr}\rho^{M}$ for any $M\geq1$, we are able to evaluate the von Neumann entropy using the following relation:
\begin{equation} \label{eq. replicat}
S^{\left(A\right)}  =  \lim_{M\to1}\frac{d}{dM}S_{M}^{\left(A\right)}
  =  \lim_{M\to1}\frac{d}{dM}\textrm{Tr}_{A}\left(\rho_{A}\right)^{M}.
\end{equation}

One can see that there is no need to take the logarithm of $\textrm{Tr}_{A}\left(\rho_{A}\right)^{M}$. This is only a mathematical simplification in the vicinity of $M \to 1$, i.e.,~when we want to reproduce von Neumann entropy by analytically continuing the derivative of the Renyi entropy. Otherwise, the presence of the logarithm is essential for the definition of the Renyi entropy.  It might be useful to further comment that the Renyi entropy without the logarithm has many names such as Tsallis entropy or power entropy, etc. However, the~presence of the logarithm is necessary for what we call the Renyi entropy. Otherwise,~we would have $\lim_{M\to 1} Tr \rho^M =1$, which, in~this important limit, cannot be a true measure of information.   

However, calculating $\textrm{Tr}_{A}\left(\rho_{A}\right)^{M}$ for a real or complex number $M$ is a hopeless task. The~`replica trick'
does the following: compute $\textrm{Tr}_{A}\left(\rho_{A}\right)^{M}$ only for integer $M$ and then analytically continue it to a general real or even complex~number.

\subsection{Time Evolution of~Entropy}
Let us mention that we limit our analysis here only to weak coupling. In~this regime, the dynamics of a quantum system are reversible and can be formulated in terms of the density matrix evolution. This time evolution depends on the the time-dependent Hamiltonian $ H \left( t \right) = H_{A} + H_{B} + H_{AB}$ as~follows:
\begin{equation}
	\label{eq. densmat evol}
	\frac{d\rho}{dt} = 
		\frac{i}{\hbar} \left[ H (t), \rho (t) \right].
\end{equation}

We transform the basis to the interaction frame by using defining a unitary operator with the non-interaction part of the Hamiltonian $U(t) =  \exp \left[ -i \left( H_{A} + H_{B} \right) t \right]$. The~density matrix transforms as $R \left( t \right) = U \left( t \right) \rho \left( t \right) U^{\dagger} \left( t \right)$, thereby not changing its entropy, neither in parts nor in total. 
In the new basis, Equation~(\ref{eq. densmat evol}) becomes
\begin{equation} 
	\label{eq. densmat evol int}
	\frac{dR}{dt} = 
		\frac{i}{\hbar} \left[ 
			U^\dagger (t) H_{AB} (t) U(t), R(t). 
		\right]
\end{equation} 

Let us refer to the interaction Hamiltonian $H_{AB}$ in the new basis as $H_{I}$, i.e.,~$H_{I} \equiv  U^{\dagger} \left( t \right) H_{AB} \left( t \right) U \left( t \right)$. The~solution to the time evolution Equation~(\ref{eq. densmat evol int}) can be written as
\begin{eqnarray} \nonumber \label{eq. solution}
R\left(t\right) & = & R_{0}+R^{\left(1\right)} + \mathcal{O}(2) \\ \nonumber
R_{0} & \equiv & R\left(0\right) \quad\textrm{noninteracting}\\
R^{\left(1\right)} & \equiv & i \int_{0}^{t}ds\left[H_{I}\left(s\right),R\left(s\right)\right]\quad\textrm{1st order}
\end{eqnarray}
This solution can be inserted back in the right side of Eq. (\ref{eq. densmat evol int}), which declares its cycle of internal interaction and we truncate the series at the second order:
\begin{eqnarray} \nonumber \label{eq. cycle}
\frac{dR\left(t\right)}{dt} & = & \Delta^{\left(1\right)}+\Delta^{\left(2\right)} + \mathcal{O}(3)\\ \nonumber
\Delta^{\left(1\right)} & \equiv & i \left[H_{I}\left(t\right),R_0\right]\quad\textrm{1st order}\\ 
\Delta^{\left(2\right)} & \equiv & - \int_{0}^{t}ds\left[H_{I}\left(t\right),\left[H_{I}\left(s\right),R\left(s\right)\right]\right]\quad\textrm{2nd order}
\end{eqnarray}

In order to find the time evolution of the Renyi and von Neumann entropies, we first notice that the unitary transformation $U(t)$, defining the basis change, also transforms any power of the density matrix, i.e.,~ \begin{equation}
R\left(t\right)^{M}=U(t)\left(\rho\left(t\right)^{M}\right)U^{\dagger}(t).
\end{equation}

Now, all we need to do is to generalize the evolution of density matrix to the powers of density matrix $\left(R\left(t\right)\right)^{M}$.
We follow the terminology of Nazarov in~\cite{Nazarov11} and name each copy of replica  $R\left(t\right)$ in the matrix $(R\left(t\right))^M$  a `world', thus $\left(R\left(t\right)\right)^{M}$ is the generalized density matrix of $M$ worlds:
\begin{eqnarray*}
	\frac{d}{dt} \left( R\left(t\right)^{M} \right) & = & \left[\frac{d}{dt}R\left(t\right)\right]\left(R\left(t\right)\right)^{M-1}+R\left(t\right)\left[\frac{d}{dt}R\left(t\right)\right]\left(R\left(t\right)\right)^{M-2}\\
	&  & +\cdots+\left(R\left(t\right)\right)^{M-2}\left[\frac{d}{dt}R\left(t\right)\right]R\left(t\right)+\left(R\left(t\right)\right)^{M-1}\left[\frac{d}{dt}R\left(t\right)\right].
\end{eqnarray*}

By substituting the solutions of Equations (\ref{eq. solution R}) to  (\ref{eq. cycle 2nd}), and~limiting the result to second order, we find the the following time evolution of the $M$-world density matrix:
\begin{eqnarray} \label{eq. evol M w} \nonumber
\frac{d}{dt}\left(R\left(t\right)^{M}\right) & = & \Delta^{\left(2\right)}R_{0}^{M-1}+R_{0}\Delta^{\left(2\right)}R_{0}^{M-2}+\cdots+R_{0}^{M-1}\Delta^{\left(2\right)}\\ \nonumber
&  & +\Delta^{\left(1\right)}\left\{ R^{\left(1\right)}R_{0}^{M-2}+R_{0}R^{\left(1\right)}R_{0}^{M-3}+\cdots+R_{0}^{M-2}R^{\left(1\right)}\right\} \\ \nonumber
&  & +R_{0}\Delta^{\left(1\right)}\left\{ R^{\left(1\right)}R_{0}^{M-3}+R_{0}R^{\left(1\right)}R_{0}^{M-4}+\cdots+R_{0}^{M-3}R^{\left(1\right)}\right\} \\ \nonumber
&  & +R_{0}^{2}\Delta^{\left(1\right)}\left\{ R^{\left(1\right)}R_{0}^{M-4}+R_{0}R^{\left(1\right)}R_{0}^{M-5}+\cdots+R_{0}^{M-4}R^{\left(1\right)}\right\} \\ \nonumber
&  & +\cdots\\ 
&  & +\left\{ R^{\left(1\right)}R_{0}^{M-2}+R_{0}R^{\left(1\right)}R_{0}^{M-3}+\cdots+R_{0}^{M-2}R^{\left(1\right)}\right\} \Delta^{\left(1\right)}.
\end{eqnarray}

This is how the $M$-world density matrix evolves in time. The~first line in Equation~(\ref{eq. evol M w}) denotes the case where the 2nd order perturbation takes place in one world while the $M - 1$ remaining worlds are left non-interacting. All these remaining terms have in common that they don't contain a 2nd order term occurring in a single replica. Instead, these terms contain two 1st order interactions, each acting in a single replica, which together combine to give a 2nd order perturbation term. These new terms have recently been found~\cite{AN14}. 

If you decide to consider higher perturbative orders, say up to $k$-th order with $k \leq M$, there will be terms like $R_{0}^{M-1} \Delta^{ (k)}$ in the expansions that have $k$ interactions taking place in one replica, leaving $M-1$ replicas noninteracting as well as terms having $k$ first-order configurations combining to give a $k$th order interaction term, such as $R_{0}^{M-k}\left(\Delta^{\left(1\right)}\right)^{k}$.  In~the case $k>M$, some of the lowest-order interactions will obviously become excluded from the~summations.

Let us show the time evolution pictorially using the following diagrams,
in which the evolution of $\left(R(t)\right)^{M}$ is shown by $M$
parallel lines, each one denoting the time evolution of one world, starting in the past at the bottom and arriving at the present time on the top. In~the following diagrams, we show five time-slices by horizontal dashed lines. Blue dots denote the interaction $H_I(t)$ and our diagrams are limited to the 2nd order only. Curly photon-like lines connect the two interactions and represent the correlation~function.

The first line of Equation~(\ref{eq. evol M w}) contains all terms that have two interactions in a single world. These~two interactions within the same world are called `self-replica interactions'. They can be illustrated pictorially by the following  diagrams in Fig. (\ref{fig1}) from left to right:
\begin{figure}[H]
\begin{centering}
\includegraphics[scale=0.6]{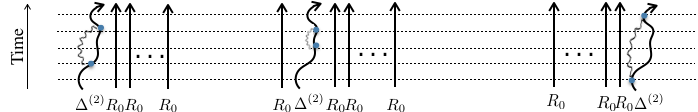}
\par\end{centering}
\caption{\bf{Diagrammatic representation of terms in the first line in Equation~(\ref{eq. evol M w}).}}
\label{fig1}
\end{figure}

\begin{figure}[H]
\begin{centering}
\includegraphics[scale=0.7]{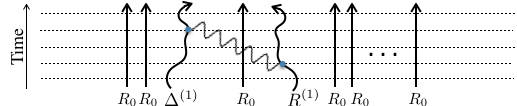}
\par\end{centering}
\caption{\bf{A typical diagram with two first order interactions acting on two different worlds.}}
\label{fig2}
\end{figure}

The following diagram in Fig. (\ref{fig2}) illustrates the typical term $\left(R_{0}\right)^{2}\Delta^{\left(1\right)}R_{0}R^{\left(1\right)}\left(R_{0}\right)^{M-4}$ from Equation~(\ref{eq. evol M w}) and pictorially shows the contribution of two first order interactions in two different worlds that together evolve the generalized density matrix of $M$ worlds in the second~order.

A typical higher order digram limited to two-correlation interactions
can diagrammatically be shown as~below in Fig. (\ref{fig3}) .

\begin{figure}[H]
\begin{centering}
\includegraphics[scale=0.7]{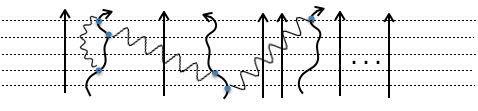}
\par\end{centering}
\caption{\bf{A typical higher order diagram.}}
\label{fig3}
\end{figure}

\subsection{Extended Keldysh~Diagrams}

In all the above diagrams, quantum states have been represented as labels on the contours.  By~definition, we know that the density matrix contains both ket and bra states. The~second order interactions can, in~fact, only take place either between two kets, two bras, or~between a ket and a bra. This internal degree of freedom makes it necessary 
to add more details to our diagrams and represent each replica with the well-known Keldysh contour diagrams~\cite{Keldysh}.
The Keldysh technique permits a natural formulation of the density matrix
dynamics in terms of path integrals, which is a generalization of
the Feynman--Vernon~formalism.

Considering that the time evolution of a quantum system takes place
by the Hamiltonian $H$, kets evolve as $|\psi\left(t\right)\rangle=\exp\left(iHt\right)|\psi\left(0\right)\rangle$
and bras evolve with the opposite phase:~$\langle\psi\left(t\right)|=\langle\psi\left(0\right)|\exp\left(iHt\right)$.
Based on this simple observation, bras (kets) evolve in the opposite (same) direction of time along the Keldysh~contour. 

The evolution of the density matrix $R$ from the initial time to the present time can diagrammatically be represented in the following way: one can start at a bra at the present time, move down along the contour to the initial time, pass there through the initial density matrix thereby changing from a bra to a ket, and~finally move upwards to end with a ket at the present time. Taking a trace from the density matrix can be shown diagrammatically by closing the contours at the present time: i.e.,~we connect the present ket to the present bra. It is of course awkward to do this for the total density matrix, as~this will simply yield one at any time; however, taking a trace is meaningful for multiple interacting~subsystems.

\begin{figure}[H]
\centering
\includegraphics[scale=0.7]{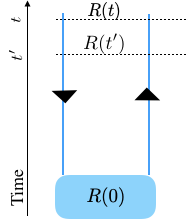} 
\includegraphics[scale=0.7]{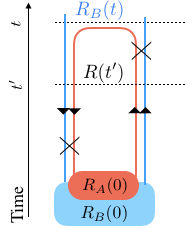}
\caption{\bf{The Keldysh diagram for the time evolution of: (\textbf{left}) one world made of one subsystem, (\textbf{right}) a world made of two interacting subsystems. Each contour represents a subsystem and the crosses denote interactions.}}
\label{fig. contours}
\end{figure}

The two subsystems $A$ and $B$ each require a contour, resulting in a double contour. We assume separability of $A$ and $B$ at the initial time:
$R\left(0\right)=R_{A}\left(0\right)R_{B}\left(0\right)$. Interaction results in energy exchange, which we represent by a cross between the two contours, somewhere between initial and present times, i.e.,~$0<t'<t$. In~the case we are interested in the evolution of one of the subsystems, say $B$, the~partial trace over $A$ should be taken, which in the diagram can be done by connecting the present bra and ket of system $A$, see the right diagram in Figure~\ref{fig. contours}. Further details about this Keldysh representation of quantum dynamics can be found in~\cite{ANS16}.

In order to evaluate the time evolution of the von Neumann and Renyi entropies, we need extended Keldysh contours in multiple parallel worlds (replicas). For~this purpose, we consider multiple copies of the Keldysh diagram, one for each world, and~add the initial state of the density matrix in each world along the contour at the initial time. The~overall trace will get the contours of different worlds~connected.

In the second order, one can find:
\begin{eqnarray} \label{eq. R ent flow B} \nonumber 
\frac{d}{dt}S_{M}^{\left(B\right)} & = & -\frac{1}{S_{M}^{\left(B\right)}}\textrm{Tr}_{B}\left\{ \Delta_{B}^{\left(2\right)}R_{B}\left(0\right)^{M-1}+R_{0}\Delta_{B}^{\left(2\right)}R_{B}\left(0\right)^{M-2}+\cdots+R_{B}\left(0\right)^{M-1}\Delta_{B}^{\left(2\right)}\right\} \\
 &  & -\frac{1}{S_{M}^{\left(B\right)}}\textrm{Tr}_{B}\left\{ \Delta_{B}^{\left(1\right)}\left[R_{B}^{\left(1\right)}R_{B}\left(0\right)^{M-2}+\cdots+R_{B}\left(0\right)^{M-2}R_{B}^{\left(1\right)}\right]\right.\nonumber \\
 &  & \qquad\qquad+R_{B}\left(0\right)\Delta_{B}^{\left(1\right)}\left[R_{B}^{\left(1\right)}R_{B}\left(0\right)^{M-3}+\cdots+R_{B}\left(0\right)^{M-3}R_{B}^{\left(1\right)}\right]+\cdots\nonumber \\
 &  & \left.\qquad\qquad+\left[R_{B}^{\left(1\right)}R_{B}\left(0\right)^{M-21}+\cdots+R_{B}\left(0\right)^{M-2}R_{B}^{\left(1\right)}\right]\Delta_{B}^{\left(1\right)}\right\}.  
\end{eqnarray}

\begin{figure}[H] 
\begin{centering} 
\includegraphics[scale=0.6]{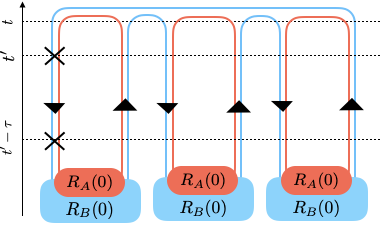}
\par\end{centering}
\caption{\bf{A diagram with two energy exchanges in one replica and no interaction in others.}}
\label{fig. ext 2nd order}
\end{figure}

The first line contains terms with second-order interactions taking place in only one world. A~typical such diagram for $M=3$ has been shown in Figure~\ref{fig. ext 2nd order}. 

\begin{figure}[H]
\begin{centering}
\includegraphics[scale=0.6]{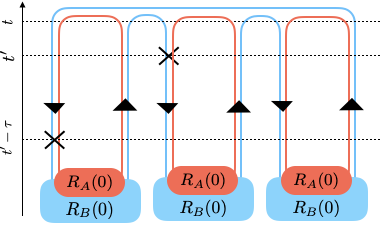}
\par\end{centering}
\caption{\bf{A  diagram with two replicas taking over 1st order interactions and the others remain intact.}}
\label{fig. ext 2nd order2}
\end{figure}

The rest of the lines other than
the first line in Equation~(\ref{eq. R ent flow B})  denote maximally no more than first-order interaction in a replica. The~diagram in Figure~\ref{fig. ext 2nd order2} shows a typical such~term.  

\subsection{Calculating the~Diagrams}

The main reason why the time evolution of entropy in Equation~(\ref{eq. R ent flow B})
has been diagrammatically represented is that, due to the multiplicity in time ordering interactions, these extended Keldysh diagrams can help to correctly determine all possible symmetries that may simplify the problem. We~need to express all `single-world' interactions that carry the highest order perturbation as well as all `cross-world' terms with lower orders of~perturbation.

We assume the interaction Hamiltonian does not implicitly depend on time through its parameters; instead, the~time dependence is globally assigned in the rotating frame and
state evolutions. The~explicit formulation of quantum dynamics and keeping track of symmetries between different diagrams have resulted in the following rules for the evaluations of the~diagrams: 
\begin{enumerate}[leftmargin=*,labelsep=5mm]
	\item 
		With each system having its own contours in each world, label each separate segment of these contours, according to the state of the associated bra or ket of that segment. The~state of the bras and kets change after an interaction, at~the initial time and at the final~time.
	
	\item 
		Starting from the present time in any of the worlds, say the leftmost world, and~encompassing the contours, the~following operators or changes must be added along the~contour:
	
	\begin{enumerate}[leftmargin=2.5em,labelsep=5mm]
		\item 
			Every interaction on a ket contour will be $\left( i / \hbar \right) H_{I} \left( t' \right)$ and will be $\left(-i/\hbar\right)H_{I}\left(t'\right)$ on a bra~contour.
		
		\item
			After passing an interaction, the~states must change. The~new states remain the same until a new interaction is encountered, or~if the initial time or the final time is~reached.
		
		\item
			A contour arriving at the initial time will capture the initial density matrix in the interaction picture $R_0$.
	\end{enumerate}
	
	\item 
		In general, the result should be integrated over the individual interaction times, i.e.,~$\int_0^\infty \int_0^\infty dt_1 dt_2 $, subject to time order between them. This can be simplified for a small quantum system coupled to a large reservoir kept at a fixed temperature. The~reason being that the correlation function of absorption and decay of particles only depends on the time difference between the two interactions~\cite{QT}. In~this case, the~double integral over $dt_1$ and $dt_2$ can be be simplified to only contain a single integral over the time difference between the two interactions, i.e.,~$\int_0^\infty d\tau $. 
\end{enumerate}

\subsection{Quantum Entropy~Production}

Let us consider that two large heat reservoirs $A$ and $B$, each
one containing many degrees of freedom and kept at a temperature, are coupled to one another via only a few
numbers of shared degrees of freedom. The~Hamiltonian can be
written as $H=H_{A}+H_{B}+H_{AB}$ with $H_{AB}$ representing the
coupled degrees of~freedom.

In order to compute the flow of a quantity between $A$ and $B$, that quantity should be conserved in the combined system $A+B$. As~we discussed in the first section of this paper, Renyi entropy is a conserved quantity in a closed system, therefore $d\ln S_{M}^{(A+B)}/dt=0$. However, one should notice that there is a difference between the conservation of physical quantities such as energy and the conservation of entropy. 
Because physical quantities linearly depend on the density matrix, 
when it is conserved for a closed system, internally it can flow from a subsystem to another one such that its production in a subsystem is exactly equal to the negative sign of its removal from the other subsystem.  However, entropy is not so. In~fact, due to nonlinear dependence of entropy on the density matrix, when it is conserved for a bipartite closed system, it is not equally added and subtracted from the subsystem due to the non-equality in Equation~(\ref{eq. noneq par flow}).

Below, we will present some example systems with rather general Hamiltonians and, using the diagram rules, we evaluated all entropy production~diagrams.

\subsubsection{Example 1: Entropy in a Two-Level Quantum Heat~Engine}

\begin{figure}[H]
\centering
\includegraphics[width=9 cm]{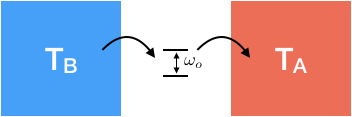}
\caption{\bf{A two-level system quantum heat bath.}}
\label{fig. tls}
\end{figure}

In Ref.~\cite{AN15}, we used the extended Keldysh technique and evaluated entropy flow for the simplest quantum heat engine in which a two-level system couples two heat baths kept at different temperatures, see Fig. (\ref{fig. tls}). After~taking all physical and informational correlations into account, we found that the exact evaluation in the second order is much different from what physical correlations predict.  Here,~we reproduce the exact result by giving a pedagogical use of the diagram evaluation described~above.

Let us consider two heat baths that are kept at different temperatures weakly interact by exchanging
the quantum energy $ \omega_{o}$. Such a quantum system can be thought of as a two-level system that couples
the two heat baths through shared excitations and de-excitations. The~Hilbert space of the two-level system contains the states $\left|0\right\rangle $ and $\left|1\right\rangle $.
The free Hamiltonian contains heat bath energy levels $E_{\alpha}^{\left(A\right)}$s and $E_{\beta}^{\left(B\right)}$s
and quantum system energies $E_{n}$ with $n=0,1$, i.e.,~$H_{0}=\sum_{\alpha}E_{\alpha}^{\left(A\right)}|\alpha\rangle\langle\alpha|+\sum_{\beta}E_{\beta}^{\left(B\right)}|\beta\rangle\langle\beta|+\sum_{n=0,1}E_{n}|n\rangle\langle n|$.

We assume the so-called `transversal' interaction is taken into account
between $A$/$B$ and the two-level system $q$. This means that they interact via exchanging the quantum of energy $  \omega_o$. Of~course, we can generalize the discussion to longitudinal interactions in which no energy is exchanged; however, since such interactions are not of immediate interest for heat transfer in quantum heat devices. we ignore~them. 

This interaction we assume for the heat bath has the following general form: $H_{int}=\sum_{n,m=0,1}\left|n\right\rangle \left\langle m\right|\left[\hat{X}_{nm}^{(A)}\left(\omega_{0}\right)+\hat{X}_{nm}^{(B)}\left(\omega_{0}\right)\right]$
subject to $E_{m}\neq E_{n}$ and $\hat{X}_{nm}$ representing energy
absorption/decay in heat baths. The~summation in $H_{int}$ can be
generalized to an arbitrary number of heat baths interacting at shared
degrees of~freedom.

Moreover, the~entire system including the two-level system is externally
driven. The~classical heat baths are naturally not influenced effectively
by the driving field; however, the~driving can pump in and out energy
to the two-level system by the following Hamiltonian $H_{dr}= \Omega\cos(\omega_{dr}t)\left(|0\rangle\langle1|+|1\rangle\langle0|\right)$.

For simplicity, we take the Hamiltonian into
the rotating frame that makes excitation/relaxation with the frequency $\omega_{dr}$. In~this frame, the excited and ground states are transformed as follows: $|1\rangle_{R}=\exp\left(i\omega_{dr}t\right)|1\rangle$
and $|0\rangle_{R}=|0\rangle$. This will introduce the unitary transformation
$U_{R}=\exp\left(i\omega_{dr}t|1\rangle\langle1|\right)$ on the Hamiltonian,
i.e., $H_{R}=U_{R}HU_{R}^{\dagger}+i \left(\partial U_{R}/\partial t\right)U_{R}^{\dagger}$.
A few lines of simplification will result in the following Hamiltonian
in the rotating frame:

\begin{eqnarray} \nonumber \label{eq. Hamil tls}
H_{R} & \equiv & H_{0}+V_{qA}+V_{qB}+V_{AB}+V_{dr},\\ \nonumber
H_{0} & = & E_{0}|0\rangle\langle0|+\left(E_{1}- \omega_{dr}\right)|1\rangle\langle1|+\sum_{\alpha}E_{\alpha}^{\left(A\right)}|\alpha\rangle\langle\alpha|+\sum_{\alpha}E_{\alpha}^{\left(B\right)}|\alpha\rangle\langle\alpha|,\\ 
V_{qA} & = & |0\rangle\langle1|\hat{X}_{01}^{(A)}\left(t\right)e^{i\omega_{dr}t}+|1\rangle\langle0|\hat{X}_{10}^{(A)}\left(t\right)e^{-i\omega_{dr}t}\equiv\sum_{n,m=0,1(n\neq m)}|n\rangle\langle m|\hat{X}_{nm}^{(A)}(t)e^{i\omega_{dr}\eta_{nm}t},\\ \nonumber
V_{qB} & = & |0\rangle\langle1|\hat{X}_{01}^{(B)}\left(t\right)e^{i\omega_{dr}t}+|1\rangle\langle0|\hat{X}_{10}^{(B)}\left(t\right)e^{-i\omega_{dr}t}\equiv\sum_{n,m=0,1(n\neq m)}|n\rangle\langle m|\hat{X}_{nm}^{(B)}(t)e^{i\omega_{dr}\eta_{nm}t},\\ \nonumber
V_{AB} & = & 0,\ \ \ \ \ \  V_{dr}  =  \frac{ \Omega}{2}\left(|0\rangle\langle1|+|1\rangle\langle0|\right),
\end{eqnarray}
with $\eta_{01}=-\eta_{10}=1$ and $\eta_{00}=\eta_{11}=0$. Given the fact that there is no direct exchange of energy between $A$ and $B$, the~density matrix can be represented as $R=R_{qA}\otimes R_{B}+R_{A}\otimes R_{qB}$ in an interaction picture, thus determining entropy flow in the heat bath $B$ will depend on the quantum system and the heat bath $B$, although~indirectly the heat bath
A will influence the quantum system. In~general, $d\left(R_{B}\right)^{M}/dt=Tr_{q}\left\{ d\left(R_{qB}\right)^{M}/dt\right\} $. Let us recall that this quantity determines the flow of von Neumann entropy and, using Equation~(\ref{eq. replicat}), it can be simplified to $dS^{(B)}/dt=\lim_{M\to1}d\left(\textup{Tr}_{B}Tr_{q}\left\{ \left(d R_{qB}/dt\right) \left(R_{qB}\right)^{M-1}+\cdots+\left(R_{qB}\right)^{M-1}\left(dR_{qB}/dt\right)\right\} \right)/dM$.~Each~term in the sum is evaluated in the interaction picture using $d R/dt=\left(-i \right)\left[V,R\right]$.~One~can show that the external driving will cause the density matrix to evolve as $\left.dR_{nm}/dt\right\rfloor _{dr}=\left(i\Omega/2\right)\left(R_{n0}\delta_{m1}+R_{n1}\delta_{m0}-\delta_{n0}R_{1m}-\delta_{n1}R_{0m}\right)$.

\begin{figure}[H]
\centering
\includegraphics[scale=0.28]{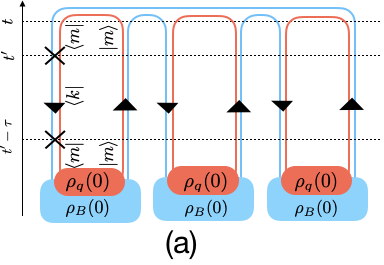}\includegraphics[scale=0.28]{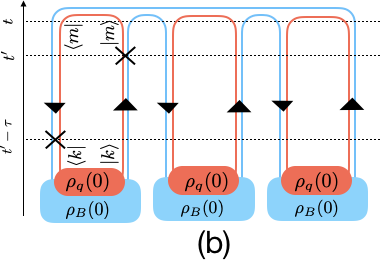}\includegraphics[scale=0.28]{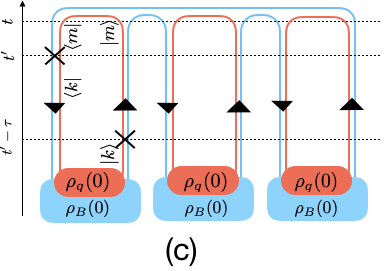}\includegraphics[scale=0.28]{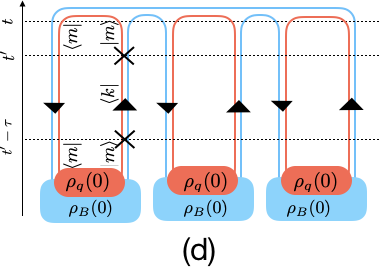}
\caption{\bf{{Self-interacting diagrams for interaction between a quantum system and a heat bath}.} }
\label{fig. o2diag}
\end{figure}

The interaction Hamiltonian evolves quantum states and below we evaluate the entropy flow in the $M=3$ example to the second order perturbation theory.  As~discussed above, there are in general two types of diagrams in the second order: (1) 'self-interacting' diagrams with second order interaction taking place in one replica, and~(2) cross-world-interacting terms in which two different replicas take on each 1st order interaction. The~self-interacting diagrams for the two-level system are listed in Figure~\ref{fig. o2diag}.

These diagrams correspond to the following flows, respectively:  

\begin{eqnarray*}
(a): &  & \frac{\left(-1\right)\int_{0}^{\infty}d\tau\textup{Tr}_{B}\left\{ \sum_{m,k=0,1(m\neq k)}\hat{X}_{mk}^{(B)}(t')\hat{X}_{km}^{(B)}\left(t'-\tau\right)\hat{ R}_{B}\hat{ R}_{mm}e^{-i\omega_{dr}\eta_{km}\tau}e^{i\omega_{dr}\left(\eta_{mk}+\eta_{km}\right)t'}\hat{ R}_{B}^{2}\right\} }{\textup{Tr}_{B}\left(\hat{ R}_{B}^{3}\right)},\\
(b): &  & \frac{\left(+1\right)\int_{0}^{\infty}d\tau\textup{Tr}_{B}\left\{ \sum_{m,k=0,1(m\neq k)}\hat{X}_{mk}^{(B)}(t'-\tau)\hat{ R}_{B}\hat{ R}_{kk}\hat{X}_{km}^{(B)}\left(t'\right)e^{-i\omega_{dr}\eta_{mk}\tau}e^{i\omega_{dr}\left(\eta_{mk}+\eta_{km}\right)t'}\hat{ R}_{B}^{2}\right\} }{\textup{Tr}_{B}\left(\hat{ R}_{B}^{3}\right)},\\
(c): &  & \frac{\left(+1\right)\int_{0}^{\infty}d\tau\textup{Tr}_{B}\left\{ \sum_{m,k=0,1(m\neq k)}\hat{X}_{mk}^{(B)}(t')\hat{ R}_{B}\hat{ R}_{kk}\hat{X}_{km}^{(B)}\left(t'-\tau\right)e^{-i\omega_{dr}\eta_{km}\tau}e^{i\omega_{dr}\left(\eta_{mk}+\eta_{km}\right)t'}\hat{ R}_{B}^{2}\right\} }{\textup{Tr}_{B}\left(\hat{ R}_{B}^{3}\right)},\\
(d): &  & \frac{\left(-1\right)\int_{0}^{\infty}d\tau\textup{Tr}_{B}\left\{ \sum_{m,k=0,1(m\neq k)}\hat{ R}_{B}\hat{ R}_{mm}\hat{X}_{mk}^{(B)}(t'-\tau)\hat{X}_{km}^{(B)}\left(t'\right)e^{-i\omega_{dr}\eta_{mk}\tau}e^{i\omega_{dr}\left(\eta_{mk}+\eta_{km}\right)t'}\hat{ R}_{B}^{2}\right\} }{\textup{Tr}_{B}\left(\hat{ R}_{B}^{3}\right)}.
\end{eqnarray*}

In all these terms, there is a time dependent factor $e^{i\omega_{dr}\left(\eta_{mk}+\eta_{km}\right)t'}$
which is identical to 1 because we always have the following relation valid: $\eta_{mk}=-\eta_{km}$.
We assume that heat baths are large and, at equilibrium, therefore the
correlation function is the same at all times $t'$ and only depends
on the time difference $\tau$ between the creation and annihilation
of a photon. In~the heat bath B, the~equilibrium correlation is defined
as $S_{mn,pq}^{\left(B\right)}\left(\tau\right)\equiv\textup{Tr}_{B}\left(\hat{X}_{mn}^{\left(B\right)}\left(0\right)\hat{X}_{pq}^{\left(B\right)}\left(\tau\right) R_{B}\right)$.
The Fourier transformation  of the correlation defines the following frequency-dependent
correlation: $S_{mn,pq}^{\left(B\right)}\left(\omega\right)=\int_{-\infty}^{\infty}d\tau\textup{Tr}_{B}\left(\hat{X}_{mn}^{\left(B\right)}\left(0\right)\hat{X}_{pq}^{\left(B\right)}\left(\tau\right) R_{B}\right)\exp\left(i\omega\tau\right)$.
Therefore, in the case of $M=1$ (i.e., the absence of the last term
$ R_{B}^{2}$), the diagrams a--d can be rewritten in terms of $S_{mn,pq}^{\left(B\right)}\left(\omega\right)$.
For example, the diagram (a) for the case of $M=1$ can be simplified
to $-\sum_{m,k=0,1(m\neq k)}\hat{ R}_{mm}\int_{0}^{\infty}d\tau\textup{Tr}_{B}\left\{ \hat{X}_{mk}^{(B)}(0)\hat{X}_{km}^{(B)}\left(\tau\right)\hat{ R}_{B}e^{-i\omega_{dr}\eta_{km}\tau}\right\} $
in which the integral is half of the domain in Fourier transformation
and therefore it can be proved to simplify to $-\sum_{m,k=0,1(m\neq k)}\hat{ R}_{mm}\left[\left(1/2\right)S_{mk,km}^{\left(B\right)}\left(\omega_{dr}\eta_{mk}\right)+i\Pi_{mk,km}\left(\omega_{dr}\eta_{mk}\right)\right]$
with $\Pi_{mn,pq}\equiv\left(i/2\pi\right)\int d\nu S_{mn,pq}^{\left(B\right)}\left(\nu\right)/\left(\omega-\nu\right)$.
What is left to be determined is the frequency-dependent correlation
function $S_{mn,pq}^{\left(B\right)}\left(\omega\right)$, which turns
out to become completely characterized by the set of reduced frequency-dependent
susceptibilities defined as $\tilde{\chi}_{mn,pq}^{\left(B\right)}\left(\omega\right)\equiv\left(\chi_{mn,pq}^{\left(B\right)}\left(\omega\right)-\chi_{pq,mn}^{\left(B\right)}\left(-\omega\right)\right)/i,$
with the dynamical susceptibility in the environment being $\chi_{mn,pq}^{\left(B\right)}\left(\omega\right)\equiv\left(-i\right)\int_{-\infty}^{0}\textup{Tr}_{B}\left\{ \left[\hat{X}_{mn}^{\left(B\right)}\left(\tau\right),\hat{X}_{pq}^{\left(B\right)}\left(0\right)\right] R_{B}\right\} \exp\left(-i\omega\tau\right)$.
The fluctuation--dissipation theorem provides a link between the equilibrium
correlation and the reduced dynamical susceptibility in the classical
thermal bath $B$ at temperature $T_{B}$. This relation is usually called
the Kubo--Martin--Scwinger (KMS) relation: $S_{mn,pq}^{\left(B\right)}\left(\omega\right)=n_{B}\left(\omega/T_{B}\right)\tilde{\chi}_{mn,pq}^{\left(B\right)}\left(\omega\right)$
with $n_{B}\left(\omega/T_{B}\right)=1/\left(\exp\left(\omega T_{B}\right)-1\right)$
being the Bose distribution and $k_{B}$ the Boltzmann~constant.

$\blacktriangleright$ \textbf{{Generalized KMS}} 

In the presence of replicas, similarly, the generalized correlations are defined. For~the case in which there are $M$ replicas in total and between creation and annihilations there are $N$ replicas with $0\leq N\leq M$,
the~generalized correlation function is defined as 
\begin{equation} 
S_{mn,pq}^{N,M\:\left(B\right)}\left(\tau\right)\equiv\frac{\textup{Tr}_{B}\left(\hat{X}_{mn}^{\left(B\right)}\left(0\right)\hat{ R}_{B}^{N}\hat{X}_{pq}^{\left(B\right)}\left(\tau\right)\hat{ R}_{B}^{M-N}\right)}{\textup{Tr}_{B}\left(\hat{ R}_{B}^{M}\right)}.
\end{equation} 

Similarly, one can show that
\begin{equation}
\frac{\int_{0}^{\infty}d\tau\textup{Tr}_{B}\left\{ \hat{X}_{mn}^{(B)}(0)\hat{ R}_{B}^{N}\hat{X}_{pq}^{(B)}\left(\tau\right)\hat{ R}_{B}^{M-N}e^{i\omega\tau}\right\} }{\textup{Tr}_{B}\left(\hat{ R}_{B}^{M}\right)}=\frac{S_{mn,pq}^{N,M\:\left(B\right)}\left(\omega\right)}{2}+i\Pi_{mn,pq}^{N,M\:\left(B\right)}\left(\omega\right),
\end{equation} 
with the definition $\Pi_{mn,pq}^{N,M\:\left(B\right)}\left(\omega\right)\equiv\left(i/2\pi\right)\int d\nu S_{mn,pq}^{N,M\:\left(B\right)}\left(\nu\right)/\left(\omega-\nu\right)$. One can also check from definitions that, for any heat bath, the following identities: $S_{mn,pq}^{N,M}\left(-\omega\right)=S_{pq,mn}^{M-N,M}\left(\omega\right)$, $\Pi_{mn,pq}^{N,M}\left(-\omega\right)=-\Pi_{pq,mn}^{M-N,M}\left(\omega\right)$,
and $\tilde{\chi}_{mn,pq}\left(-\omega\right)=-\tilde{\chi}_{pq,mn}\left(\omega\right)$.

Fourier transformation of this generalized correlation will define the frequency-dependent generalized correlation and, following the same mathematics as above, one can show at equilibrium thermal bath of temperature $T_{B}$ that all correlation functions can be determined through a generalized KMS relation:
\begin{equation} \label{eq. gen kms}
S_{mn,pq}^{N,M\:\left(B\right)}\left(\omega\right)=n_{B}\left(\frac{\omega}{T_{B}}\right)\tilde{\chi}_{mn,pq}^{\left(B\right)}\left(\omega\right)e^{N\frac{\omega}{k_{B}T_{B}}}.
\end{equation}

Further details can be found in~\cite{AN14}. 
$\blacktriangleleft$

Using these definitions as well as Equation~(\ref{eq. gen kms}), the~sum of diagrams (a)--(d) in Figure~\ref{fig. o2diag} can be further simplified to
\begin{eqnarray} \nonumber 
 &  & \sum_{m,k=0,1(m\neq k)}\hat{ R}_{mm}\left\{ -\left(\frac{1}{2}S_{km,mk}^{3,3\:\left(B\right)}\left(\omega_{dr}\eta_{mk}\right)+i\Pi_{km,mk}^{3,3\:\left(B\right)}\left(\omega_{dr}\eta_{mk}\right)\right)\right.\\ \nonumber 
  &  & \qquad\qquad\qquad\qquad \left. -\left(\frac{1}{2}S_{mk,km}^{0,3\:\left(B\right)}\left(\omega_{dr}\eta_{km}\right)+i\Pi_{mk,km}^{0,3\:\left(B\right)}\left(\omega_{dr}\eta_{km}\right)\right)\right\}, \\ \nonumber 
 &  & \sum_{m,k=0,1(m\neq k)}\hat{ R}_{kk} \left\{ + \left(\frac{1}{2}S_{mk,km}^{1,3\:\left(B\right)}\left(\omega_{dr}\eta_{km}\right)+i\Pi_{mk,km}^{1,3\:\left(B\right)}\left(\omega_{dr}\eta_{km}\right)\right) \right. \\ \nonumber 
 &  & \qquad\qquad\qquad\qquad \left. +\left(\frac{1}{2}S_{km,mk}^{2,3\:\left(B\right)}\left(\omega_{dr}\eta_{mk}\right)+i\Pi_{km,mk}^{2,3\:\left(B\right)}\left(\omega_{dr}\eta_{mk}\right)\right) \right\}, \\ 
 & = & \sum_{m,k=0,1(m\neq k)} -S_{mk,km}^{0,3\:\left(B\right)}\left(\omega_{dr}\eta_{km}\right) \hat{ R}_{mm} +S_{mk,km}^{1,3\:\left(B\right)}\left(\omega_{dr}\eta_{mk}\right) \hat{ R}_{kk}.  \label{eq. semi ent}
\end{eqnarray}

In total, there are $M$ number of terms similar to the last line in Equation~(\ref{eq. semi ent}) associated with similar diagrams at $M$ worlds. It is important to notice that these self-replica correlated terms are determined in fact only by physical correlations and they make already known results for the flow of von Neumann entropy in the heat bath~\cite{Alicki}.  To~see this more in more detail, one can expand the summation and use the KMS relation and its generalized version in Equation~(\ref{eq. gen kms}). After~generalizing the result for $M$ replicas, taking derivative with respect to $M$ and analytically continuing the result to $M \to 1$, the incoherent part of flow in von Neumann entropy is
\begin{equation}
\left. \frac{dS^{(B)}}{dt}\right|_{\textup{incoherent}} = -\frac{1}{T_B} \left( \Gamma^{(B)}_{\uparrow} p_0 -  \Gamma^{(B)}_{\downarrow} p_1 \right),
\end{equation}
with $\Gamma^{(B)}_\uparrow \equiv \tilde{\chi}\left(n_{B}\left(\omega_{dr}/T_{B}\right)+1\right)$ and $\Gamma^{(B)}_\downarrow \equiv \tilde{\chi} n_{B}\left(\omega_{dr}/T_{B}\right)$,  $\tilde{\chi}\equiv \tilde{\chi}_{10,01}$, and~$p_{n}\equiv R_{nn}$. These are only self-interacting replicas, which are incomplete as they ignore the following~diagrams.

\begin{figure}[H]
\includegraphics[scale=0.28]{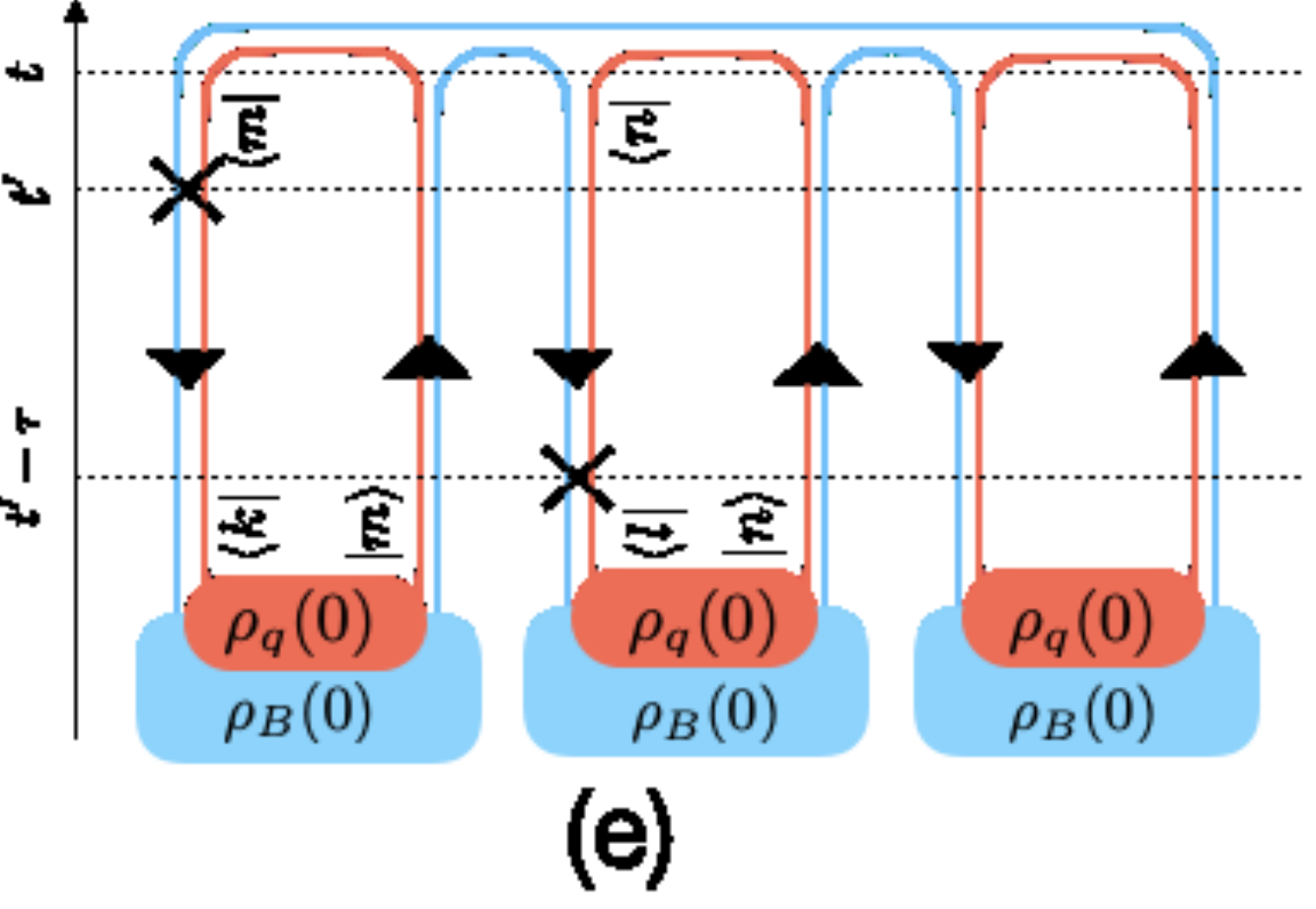}\includegraphics[scale=0.28]{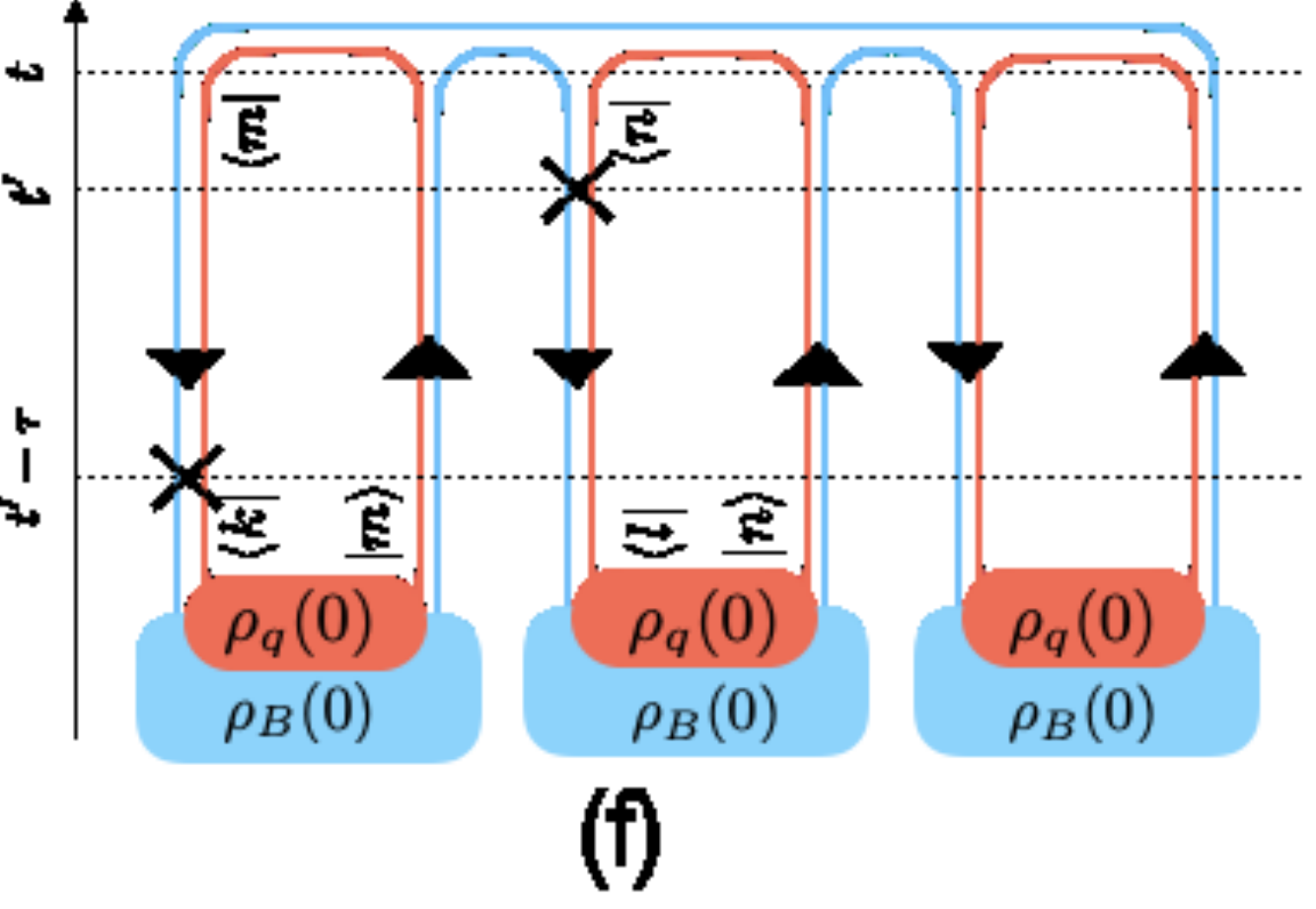}\includegraphics[scale=0.28]{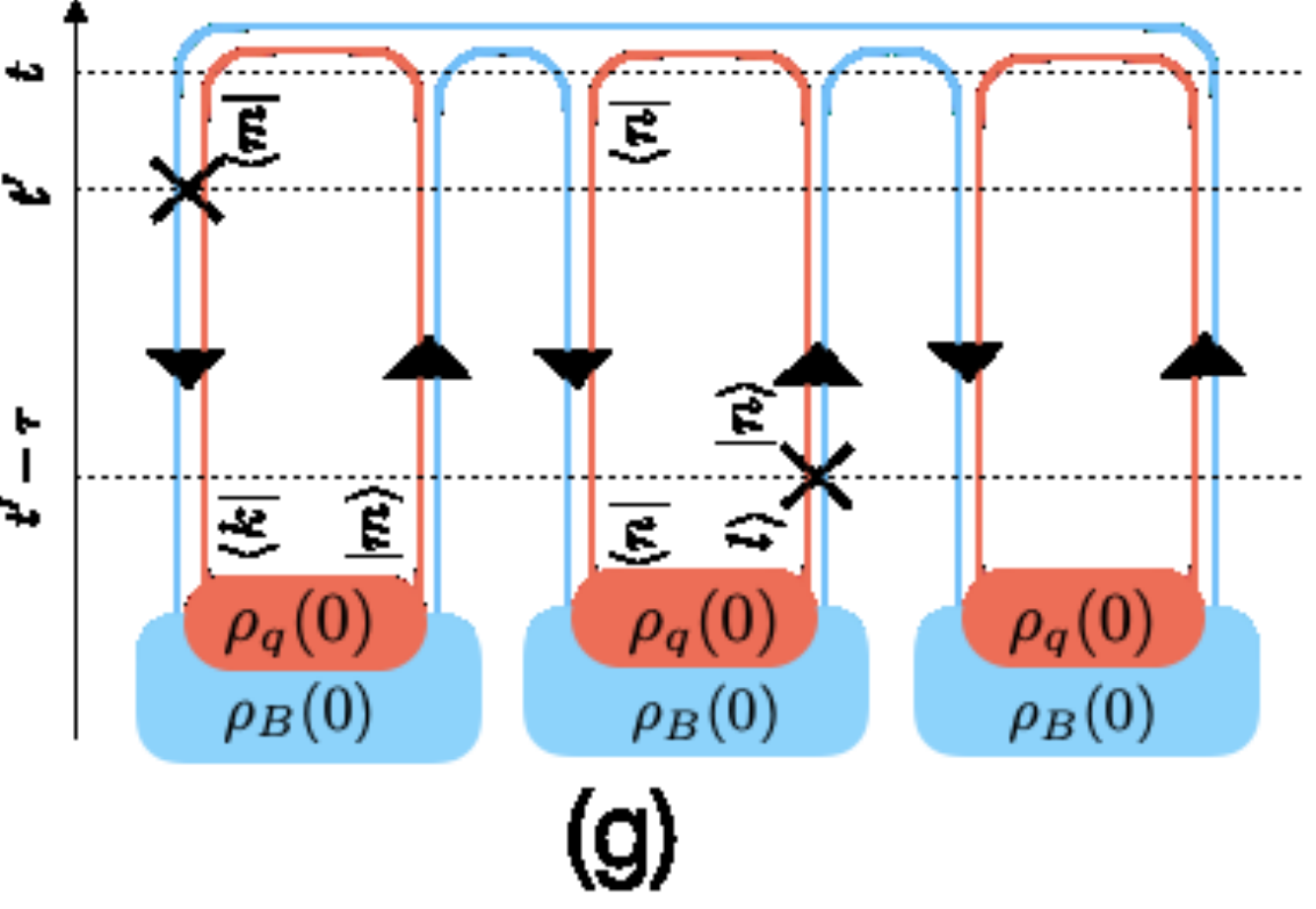}\includegraphics[scale=0.28]{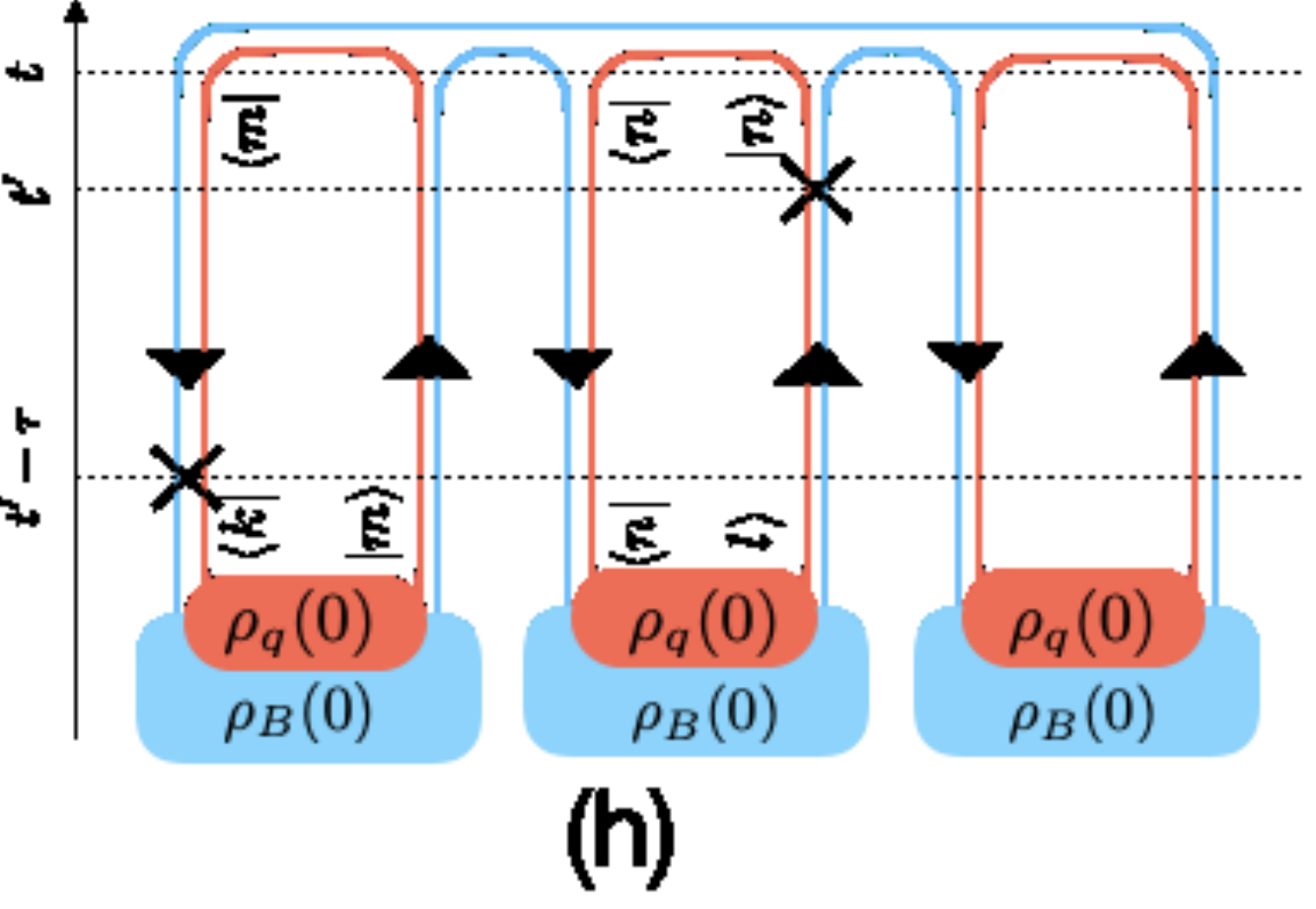} 
\includegraphics[scale=0.28]{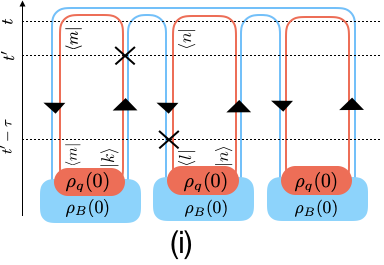}\includegraphics[scale=0.28]{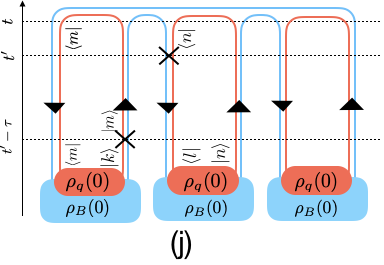}\includegraphics[scale=0.28]{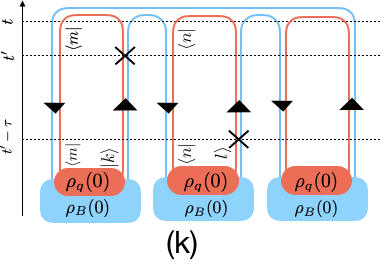}\includegraphics[scale=0.28]{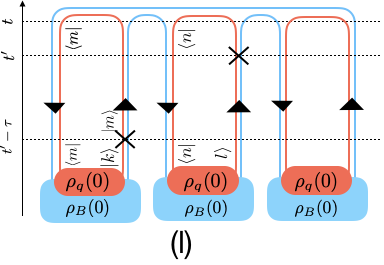}
\caption{\bf{{Cross-replica interacting diagrams for a quantum system and a heat bath}.} }
\label{fig. o2nondiag}
\end{figure}

The new diagrams are the cross-world interactions. As~discussed
previously, cross-world diagrams cannot transfer physical quantities as they rely
on the fact that entropy depends nonlinearly on the density matrix and therefore it is not a physical observable quantity. Some of these types of diagrams are shown in Figure~\ref{fig. o2nondiag}---for the case that one interaction takes place in the leftmost replica and the second interaction in the middle replica, thus leaving the third replica intact. 

\begin{eqnarray*}
(e): &  & -\int_{0}^{\infty}d\tau\textup{Tr}_{B}\left\{ \sum_{m,n,k,l}\hat{X}_{mk}^{(B)}(t')\hat{ R}_{B}\hat{ R}_{mk}\hat{X}_{nl}^{(B)}\left(t'-\tau\right)\hat{ R}_{B}\hat{ R}_{nl}e^{-i\omega_{dr}\eta_{nl}\tau}\delta_{E_{nl},E_{km}}\hat{ R}_{B}\right\} /\textup{Tr}_{B}\left(\hat{ R}_{B}^{3}\right), \\
(f): &  & -\int_{0}^{\infty}d\tau\textup{Tr}_{B}\left\{ \sum_{m,n,k,l}\hat{X}_{mk}^{(B)}(t'-\tau)\hat{ R}_{B}\hat{ R}_{mk}\hat{X}_{nl}^{(B)}\left(t'\right)\hat{ R}_{B}\hat{ R}_{nl}e^{-i\omega_{dr}\eta_{mk}\tau}\delta_{E_{mk},E_{ln}}\hat{ R}_{B}\right\} /\textup{Tr}_{B}\left(\hat{ R}_{B}^{3}\right),\\
(g): &  & \int_{0}^{\infty}d\tau\textup{Tr}_{B}\left\{ \sum_{m,n,k,l}\hat{X}_{mk}^{(B)}(t')\hat{ R}_{B}\hat{ R}_{mk}\hat{ R}_{B}\hat{ R}_{ln}\hat{X}_{ln}^{(B)}\left(t'-\tau\right)e^{-i\omega_{dr}\eta_{ln}\tau}\delta_{E_{ln},E_{km}}\hat{ R}_{B}\right\} /\textup{Tr}_{B}\left(\hat{ R}_{B}^{3}\right),\\
(h): &  & \int_{0}^{\infty}d\tau\textup{Tr}_{B}\left\{ \sum_{m,n,k,l}\hat{X}_{mk}^{(B)}(t'-\tau)\hat{ R}_{B}\hat{ R}_{mk}\hat{ R}_{B}\hat{ R}_{ln}\hat{X}_{ln}^{(B)}\left(t'\right)e^{-i\omega_{dr}\eta_{mk}\tau}\delta_{E_{mk},E_{nl}}\hat{ R}_{B}\right\} /\textup{Tr}_{B}\left(\hat{ R}_{B}^{3}\right),\\
(i): &  & \int_{0}^{\infty}d\tau\textup{Tr}_{B}\left\{ \sum_{m,n,k,l}\hat{ R}_{B}\hat{ R}_{km}\hat{X}_{km}^{(B)}(t')\hat{X}_{nl}^{(B)}\left(t'-\tau\right)\hat{ R}_{B}\hat{ R}_{nl}e^{-i\omega_{dr}\eta_{nl}\tau}\delta_{E_{nl},E_{mk}}\hat{ R}_{B}\right\} /\textup{Tr}_{B}\left(\hat{ R}_{B}^{3}\right),\\
(j): &  & \int_{0}^{\infty}d\tau\textup{Tr}_{B}\left\{ \sum_{m,n,k,l}\hat{ R}_{B}\hat{ R}_{km}\hat{X}_{km}^{(B)}(t'-\tau)\hat{X}_{nl}^{(B)}\left(t'\right)\hat{ R}_{B}\hat{ R}_{nl}e^{-i\omega_{dr}\eta_{km}\tau}\delta_{E_{km},E_{ln}}\hat{ R}_{B}\right\} /\textup{Tr}_{B}\left(\hat{ R}_{B}^{3}\right),\\
(k): &  & -\int_{0}^{\infty}d\tau\textup{Tr}_{B}\left\{ \sum_{m,n,k,l}\hat{ R}_{B}\hat{ R}_{km}\hat{X}_{km}^{(B)}(t')\hat{ R}_{B}\hat{ R}_{ln}\hat{X}_{nl}^{(B)}\left(t'-\tau\right)e^{-i\omega_{dr}\eta_{ln}\tau}\delta_{E_{ln},E_{mk}}\hat{ R}_{B}\right\} /\textup{Tr}_{B}\left(\hat{ R}_{B}^{3}\right),\\
(l): &  & -\int_{0}^{\infty}d\tau\textup{Tr}_{B}\left\{ \sum_{m,n,k,l}\hat{ R}_{B}\hat{ R}_{km}\hat{X}_{km}^{(B)}(t'-\tau)\hat{ R}_{B}\hat{ R}_{ln}\hat{X}_{ln}^{(B)}\left(t'\right)e^{-i\omega_{dr}\eta_{km}\tau}\delta_{E_{km},E_{nl}}\hat{ R}_{B}\right\} /\textup{Tr}_{B}\left(\hat{ R}_{B}^{3}\right),
\end{eqnarray*}
where we used the following identity $e^{i\omega_{dr}\left(\eta_{mn}+\eta_{pq}\right)t'}=\delta_{E_{mn},E_{qp}}$.

One can evaluate all diagrams associated with a general number of replicas
using the above example. After~carefully analyzing all diagrams and proper simplifications---see~\cite{AN14}---the flow of Renyi entropy $dS_{M}/dt$ in the heat bath $B$ can be found, and~consequently the so-called coherent part of entanglement (von Neumann) entropy can be found as follows:
\begin{equation} \label{eq. coh ent}
\left. \frac{dS^{(B)}}{dt}\right|_{\textup{coherent}} = - \frac{\Gamma^{(B)}_{\downarrow} - \Gamma^{(B)}_{\uparrow} }{T_B}   \left| R_{01}\right|^2. 
\end{equation}

This is the new part of the entropy flow that comes from the generalized KMS correlations. We~call this part the coherent part because it is nonzero for degenerate states or equivalently a two-level system driven by their detuning~frequency.  

Therefore, the entanglement entropy flow is naturally separated into two parts and therefore it is equal to the sum between the two parts:
\begin{eqnarray} \label{eq. total ent} \nonumber
\frac{dS^{(B)}}{dt} &=& \left. \frac{dS^{(B)}}{dt}\right|_{\textup{incoherent}} + \left. \frac{dS^{(B)}}{dt}\right|_{\textup{coherent}}, \\ 
&=&-\frac{1}{T_B} \left( \Gamma_{\uparrow} p_0 -  \Gamma_{\downarrow} p_1 \right)  - \frac{\Gamma_{\downarrow} - \Gamma_{\uparrow} }{T_B}   \left| R_{01}\right|^2, 
\end{eqnarray}
in which the first term on the second line is what in textbooks has so far been mistakenly taken as total entropy~flow. 

As we can see, Equation~(\ref{eq. total ent}) is not directly related to energy
flow---which here corresponds to the incoherent part instead of a finite flow that depends on the quantum coherence $(R_{01})^2$.  

\begin{figure}[H]
(a) \includegraphics[scale=0.85]{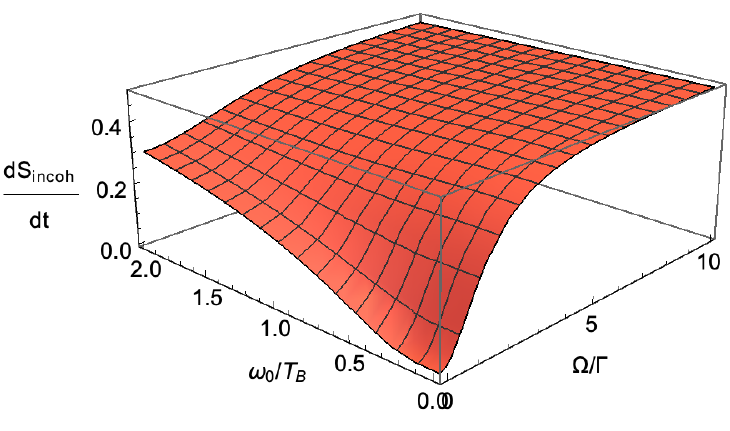}\ \ \ \  \ \ \ \  (b) \includegraphics[scale=0.85]{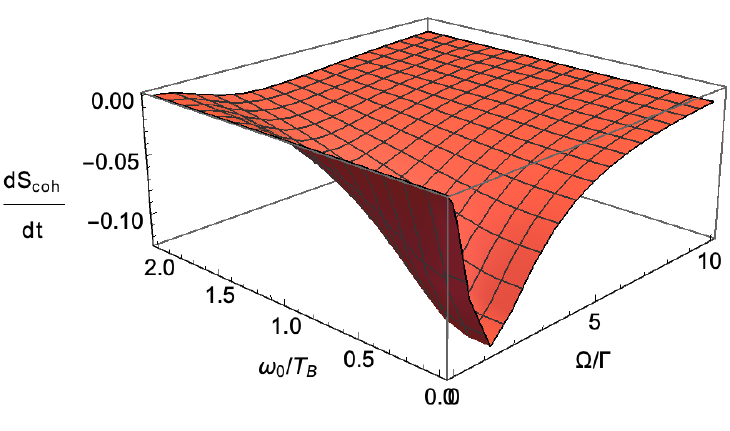} 
\caption{\bf{Entropy production in a probe bath that is kept at zero temperature and is coupled to a two-level system depicted in Figure~\ref{fig. tls}. The~entropy is the sum of two parts: the incoherent and the coherent parts.  (\textbf{a}) the incoherent part of entropy is nothing new and can be determined by standard correlations. It is positive by the convention that  entropy enters from a higher temperature bath (via the two-level system); (\textbf{b}) the coherent part of entropy is a previously unknown part as it comes from the informational correlations between different replicas. This part depends quadratically on the off diagonal density. Quite nontrivially, this part of entropy is negative and summing it with the incoherent part will result in a positive flow yet with much smaller magnitude for entropy at small driving amplitudes. }} 
\label{fig. prod}
\end{figure}

Consider that the two-level system with energy difference $ \omega_o$ is driven at the same frequency, i.e.,~$H=\Omega \cos (\omega_o t)$ and weakly coupled to two heat reservoirs at temperatures $T_A$ and $T_B$. From~Equation~(1) of Ref.~\cite{AN14}, one can find the following  time evolution equations for the density matrix and setting them to zero determines the stationary solutions:
\begin{eqnarray} \label{eq. dens eq} \nonumber
\frac{dR_{11}}{dt} &=&-\frac{i\Omega}{2}\left( R_{01} - R_{10}\right)-\Gamma_{\downarrow} R_{11}+\Gamma_{\uparrow} R_{00}=0, \\ \nonumber
\frac{dR_{01}}{dt} &=&-\frac{i\Omega}{2}\left( R_{11} - R_{00}\right)-\frac{1}{2} \left( \Gamma_{\downarrow} + \Gamma_{\uparrow} \right) R_{01}=0,  \   \  \
R_{00}+R_{11}=1,
\end{eqnarray}
which finds the stationary ground state population $R_{00}=(\gd (\gd +\gu )+\Omega^2)/((\gd + \gu)^2+2\Omega^2)$ and the stationary off-diagonal density matrix element $R_{10}=-i\Omega (1- 2 R_{00})/( \gd + \gu )$, with~$\gd\equiv \gd^{(A)}+\gd^{(B)}$ and $\gu\equiv \gu^{(A)}+\gu^{(B)}$. By~considering that$B$is a probe environment with zero temperature, substituting all solutions in Equation~(\ref{eq. coh ent}), the incoherent and coherent parts of entropy flow in the probe environment have been plotted in Figure~\ref{fig. prod} for different driving amplitudes and $\omega_0/T_A$.

\subsubsection{Example 2: Entropy in a Four-Level Quantum Photovoltaic~Cell}

Scovil and Schulz--DuBois first introduced a model of a quantum heat engine (SSDB heat engine) in which a single three-level atom, consisting of a ground and two excited states, is in contact with two heat baths~\cite{{Scovil},{SDu}}. A~large enough difference between the heat bath temperatures can create population inversion between the two excited states and a coherent light output. One hot photon is absorbed and one cold photon is emitted; therefore, a laser photon is produced. The~SSDB heat engine model gives a clear demonstration of the quantum thermodynamics. However, we notice that some detailed properties of this lasing heat engine, e.g.,~the threshold behavior and the statistics of the output light, are still not well studied.  There are a number of applications for the model, such as light-harvesting biocells, photovoltaic cells, etc. 

{Since then, the~model has been modified to describe other systems such as light-harvesting biocells, photovoltaic cells, etc.}

\begin{figure}[H]
\begin{centering} 
\includegraphics[scale=0.52]{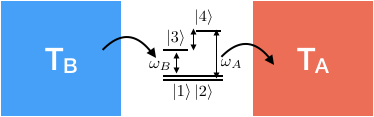}
\par \end{centering}
\caption{\bf{A four-level doubly degenerate photovoltaic cell.}}
 \label{fig. photo}
\end{figure}

Recently, in~Ref.~\cite{AN17}, one of us studied the entropy flow using the replica trick for a 4-level photovoltaic cell with two degenerate ground states and two excited states, see Figure~\ref{fig. photo}. This heat engine was first proposed by Schully in~\cite{schully13} and recently studied in many further details by Schully and others~\cite{{schully19},{mark}}. 

After finding all extended Keldysh diagrams for an arbitrary Renyi degree $M$, evaluating all self-interacting and cross-interacting diagrams and simplifying the results, the~von Neumann entropy flow in heat bath $A$ becomes~\cite{AN17}:
\begin{eqnarray} \nonumber
  \left. \frac{dS}{dt}\right|_{A} & =&  \frac{1}{T_A} \bigg\{      {\gamma}    p_4 -  \omega_A \tilde{\chi}_{42} \bar{n}\left(\frac{\omega_A}{T_A}\right) p_2   -  \omega_A \tilde{\chi}_{41} \bar{n}\left(\frac{\omega_A}{T_A}\right)  p_1  \bigg.  
\\  && \nonumber  \ \  
 -   \tilde{\chi}_{14,42} \left[ \omega_A \bar{n}\left(\frac{\omega_A}{T_A}\right) + \omega_A \bar{n}\left(\frac{\omega_A}{T_A}\right)\right] \textup{Re} R_{12}      \nonumber
 \\  &&  \bigg. - \frac{1}{2} \sum_{i=1,2}  \omega_A \tilde{\chi}_{14,42} |R_{12}|^2     \bigg\}. 
 \label{eq. rflow ex}
\end{eqnarray}

The first two lines can be found using physical correlations. The~last line, however, which plays an essential role in the entropy evaluation, can be obtained only through informational correlations. Here, the state probabilities are $p_x\equiv R_{xx}$ with $x$ being $1,2,3,4$ and depending on the characteristics of all heat baths. The~dynamical response function is $\tilde{\chi}_{\alpha i}\equiv \tilde{\chi}_{i\alpha,  \alpha i} (\omega_{i\alpha})$  with $i=1,2$ and $\alpha=3,4$, and~  $\tilde{\chi}_{1\alpha,\alpha 2}=\sqrt{\tilde{\chi}_{\alpha 1} \tilde{\chi}_{\alpha 2}}$. Moreover, ${\gamma}\equiv \sum_{i=1,2 } \left[\bar{n}\left(\omega_A/{T_A}\right)+1\right]  \omega_A \tilde{\chi}_{3i} $.  

In order to evaluate the stationary value of the entropy flow in this heat bath, we must solve the quantum master equation for the density matrix time evolution. This can be found in Ref.~\cite{AN17}. The~solution is such that the coupling between the environment and the quantum system introduces decoherence in quantum states. Energy exchange between the heat bath and a quantum system introduces a limited coherence time, namely $\tau_1$, for~quantum state probabilities. The~phase of a quantum state  can fluctuate and, depending on environmental noise, the lifetime of quantum state can be limited to $\tau_2$. These two coherence times affect all elements of the density matrix. From~solving the quantum Bloch equation, one can see that the only stationary solution in the off-diagonal part is the imaginary part of $R_{12}$ whose real part of exponential decay due to dephasing is: $\textup{Im} R_{12}\sim \exp(t/\tau_2)$.

One can substitute the stationary solution of the density matrix in Equation~(\ref{eq. rflow ex}) and the flow of entropy in the heat bath changes depending on the dephasing time---see Figure~2a,b in~\cite{AN17}. In~fact, increasing the dephasing time will increase the contribution of the coherent part of the entropy flow, i.e.,~information correlations.  This will reduce the total entropy flow in the heat bath, which will equivalently increase the output power in this photovoltaic~cell.

\subsubsection{Example 3: Entropy in a Quantum Resonator/Cavity Heat~Engine}

\begin{figure}[H]
\begin{centering}
\includegraphics[scale=0.52]{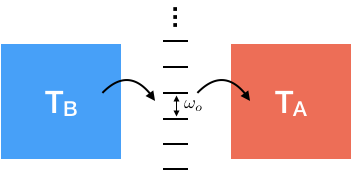}
\par \end{centering}
\caption{\bf{A quantum cavity heat engine.}}
\label{fig. harmonicos}
\end{figure}

Using a rather different technique---i.e., the correspondence between entropy and statistics of energy transfer that we discuss in the next section---in~\cite{{AN15}, {ANS16}}, we calculated entropy production for a resonator/cavity coupled two different environments kept at two different temperatures, see Fig. (\ref{fig. harmonicos}). One of the two baths is a probe environment at a temperature of zero for which we calculate the flow of~entropy.

Knowing how entropy flows as the result of interactions between the resonator, cavity and other parts of the circuit can help to obtain important information about the possibility of leakage or dephasing in the system and ultimately give rise to modifications of quantum circuits~\cite{qcom}. A~good understanding of cavities/resonators is beneficial to search for the nature of non-equilibrium quasiparticles in quantum circuits~\cite{{Catelani},{qp1}}. This can help with detecting light particles like muons whose tunnelling in a quantum circuit can signal a sudden jump in the entropy flow~\cite{{ANSquai},{ANSqs},{ANSps}}. Given that entropy flow can be measured by the full counting statistics of energy transfer, see the next section, it is important to keep track of entropy flow in a~resonator.   

Again, we use the standard technique that we described above. Let us consider a single harmonic oscillator of frequency $\omega_0$ and Hamiltonian $ \hat  H=\omega_0 (\hat{a}^\dagger \hat{a} +1/2),$ which is coupled to a number of environments at different temperatures with different coupling strengths. We concentrate on a probe environment that is weakly coupled to the oscillator. In~addition, the~oscillator is driven by an external force at frequency $\Omega$.  We calculate the Renyi flow and consequently the von Neumann entropy flow of the probe environment. The~coupling Hamiltonian between the harmonic oscillator and the probe reservoir is $\hat{H}(t)= \hat{X}(t) \hat{a}^\dagger(t) +h.c. $, with~$\hat{X}$ being the probe reservoir operator. The~Fourier transform of the correlator is: $S_{mn}(\omega)=\int \exp(-i\omega t)S_{mn}(t) d\omega/2\pi$. Due to the conservation of energy, the~energy exchange occurs either with quantum $ \Omega$ or with quantum $ \omega_0$. 

We note that the time dependence of the average of two operators can be written as $\langle \hat a^\dagger(t) \hat a(t') \rangle = \langle \langle \hat a^\dagger \hat a\rangle \rangle  e^{i\omega_0(t-t')} + \langle \hat a (t)\rangle \langle \hat a^\dagger (t')\rangle  $, where the time dependence of $\langle a (t) \rangle $  is due to the driving force and therefore oscillates at frequency $\Omega$: $\langle a(t) \rangle = \langle a \rangle_{_+} \exp({i\Omega t}) + \langle a \rangle_{_-} \exp({-i \Omega t})$. This corresponds to the fact that the oscillator can oscillate both at its own frequency and at the frequency of external~force. 

Obtaining the entropy flows from the extended Keldysh correlators is straightforward. The~generalized KMS relation in Equation~(\ref{eq. gen kms}) helps to describe the correlators in the thermal bath $B$ in terms of their dynamical susceptibility. The~result can be summarized as follows:
\begin{equation} \nonumber
\label{eq. HOi}
\frac{dS_M^{(B)}}{dt} =  \frac{M\bar{n}\left( {M\omega_0}/{T_B}\right)  \tilde{\chi}}{\bar{n}({(M-1)\omega_0}/{T_B})\ \bar{n}\left({\omega_0}/{T_B}\right)}       \left\{ \langle \langle a^\dagger a \rangle \rangle   e^{\frac { \omega_0}{ T_B}} -   \langle \langle  a a^\dagger \rangle \rangle   \right\},
  \end{equation}
where we defined $T_\textup{resonator}$  to be the effective temperature of the harmonic oscillator $\langle \langle a a^\dagger \rangle \rangle =\bar{n}(\omega_0/T_\textup{resonator})+1$ and $\langle \langle a^\dagger a \rangle \rangle =\bar{n}(\omega_0/T_\textup{resonator})$.     Taking the derivative with respect to $M$ and analytically continuing the result in the limit of $M \to 1$ will determine the thermodynamic entropy~flow:
\begin{equation} 
\label{eq. HOfinal}
 \frac{dS_M^{(B)}}{dt} =      \frac{1}{T_B}  \left\{ \bar{n}\left({\omega_0}/{T_\textup{resonator}}\right)   - \bar{n}\left({\omega_0}/{T_B}\right)       \right\}.
  \end{equation}

The entropy flow changes sign at the onset temperature $T_{resonator} = T_B$. Moreover, after~the exact evaluation of the incoherent part of the entropy flow, one should notice that it contains some terms proportional to $\langle a \rangle $ and $\langle a\dagger \rangle$. These terms oscillate with the external drive and are nonzero. However,~they are all cancelled out by the coherent part of entropy flow such that the overall flow will only depend on the temperatures, and~not on the driving force. Therefore, the entropy flow is robust in the sense that it only depends on the temperatures of the probe and harmonic oscillator and is completely insensitive to the external driving~force.

The insensitivity of entropy flow to external driving force is interesting and a direct result of including coherent flow of entropy that is absent in semi-classical analysis. The~difference can put the coherent entropy flow into an experimental~verification. 

In the absence of cross-replica correlators, the~thermodynamic entropy of a probe environment, coupled to a thermal bath via a resonator, will dramatically depend on the amplitude of the external driving. If~no such dependence on the driving amplitude is found, then this is an indication that they are absent; they are in fact eliminated by quantum coherence!

\section{Linking Information to Physics: A New~Correspondence}
As discussed above, the~Renyi entropies in quantum physics are considered
unphysical, i.e.,~non-observable quantities, due to their nonlinear dependence on the density matrix. Such~quantities cannot be determined from immediate measurements; instead, their quantification seems to be equivalent to determining the density matrix. This requires reinitialization of the density matrix between many successive measurements. Therefore, the~Renyi entropy flows between the systems are conserved measures of nonphysical quantities. An~interesting and nontrivial question is: Is there any relation between the Renyi entropy flows and the physical flows? 

An idea of such a relation was first put forward by Levitov and Klich in~\cite{LevitovKlysh}, where they proposed that entanglement entropy flow in electronic transport can be quantified from the measurement of the full counting statistics (FCS) of charge transfers~\cite{{Levitov},{Pilgrim},{LB1},{LB2}}. The~validity of this relation is restricted to zero temperature and obviously to the systems where interaction occurs by means of charge transfer. Recently, we presented a relation that is similar in spirit~\cite{AN15}. We derived a correspondence for coherent and incoherent second-order diagrams in a general time-dependent~situation. 

This relation gives an exact correspondence between the informational measure of Renyi entropy flows and physical observables, namely, the~full counting statistics of energy transfers~\cite{{Tero},{Pilgrim}}.

We consider reservoir $B$ and quantum system $q$. We assume that $B$ is infinitely large and is kept in thermal equilibrium at temperature $T_{B}$. System $q$ is arbitrary as it may carry several degrees of freedom as well as infinitely many. It does not have to be in thermal equilibrium and is in general subject to time-dependent forces. It is convenient to assume that these forces are periodic with a period of $\tau$; however, the~period does not explicitly enter the formulation of our result, which is also valid for aperiodic forces. The only requirement is that the flows of physical quantities have stationary limits. The~stationary limits are determined after averaging instant flows over a period and---for aperiodic forces---by averaging over a sufficiently long time interval. In~the case of energetic interactions, energy transfer is statistical. The~statistics can be described by the generating function of the full counting statistics (FCS), namely `FCS Keldysh~actions'.

Recently, in Ref.~\cite{AN15}, we proved that the flow of thermodynamic entropy as well as the flow of Renyi entropy between two heat baths via a quantum system is exactly equivalent to the difference between two FCS Keldysh actions of incoherent and coherent energy transfers. In~the limit of long $\tau$ and for a typical reservoir $B$ with temperature $T_{B}$, the incoherent and coherent FCS Keldysh actions are $f_{i}\left(\xi,T_{B}\right)$ and $f_{c}\left(\xi,T_{B}\right)$, with~$\xi$ being the counting
field of energy transfer. These generating functions can be determined
using Keldysh diagrams, see~\cite{ANS16}. After~their evaluation, one finds
the statistical $m$-th cumulant function $C_{m}$ by taking the derivative
of the generating function in the limit of zero counting function, i.e.,~$C_{m}=\lim_{\xi\to0}\partial^{m}f/\partial\xi^{m}$.

In fact, any physical quantity should depend on the cumulants and consequently on a zero counting field. However, informational measures are exceptional. Detailed analysis shows that the flow of Renyi entropy of degree $M$ in the reservoir $B$ at equilibrium temperature $T_{B}$ is exactly, and~unexpectedly, the~following: $dS_{M}\left(T_{B}\right)/dt=M\left[f_{i}(\xi^{*},T_{B}/M)-f_{c}(\xi^{*},T/M)\right]$
with $\xi^{*}\equiv i(M-1)/T_{B}$. Notice that in this correspondence
the temperature on the left side is $T_{B}$ while it is $T_{B}/M$
on the right side. In addition, it is important to notice that the entropy is evaluated by using the generating function of full counting statistics at nonzero counting field $\xi^{*}$. This relation is valid in the weak-coupling limit where the interaction between the systems can be treated~perturbatively.

\section{Discussion}

Currently, `time' does not play any essential role in quantum information theory. In~this sense, quantum theory is underdeveloped similarly to how quantum physics was underdeveloped before Schr\"odinger introduced his wave equation. In~this review article, we discussed a fascinating extension of the Keldysh formalism that consistently copes with the problem of time for one of the central quantities in quantum information theory: entropy.  We characterized the flows of conserved entropies (both Renyi and von Neumann entropies) and illustrated them diagrammatically to introduce new correlators that have been absent so far in the~literature.

Given that entropy is not an observable, as~it is a nonlinear function of the density matrix, one can use a probe environment to make an indirect measurement of the entropy in light of the new correspondence between entropy and full counting statistics of energy transfer. This can be done equally well for the imaginary and real values of the
characteristic parameter. The~measurement procedures may be complex,
yet they are feasible and physical. The~correspondence can have many other
advantages. For~instance, a~complete understanding of entropy flows
may help to identify the sources of fidelity loss in quantum communication and may help to develop methods to control or even prevent~them.


\end{document}